%% file: Humer.tex
\documentclass[10pt, conference, anonymous]{IEEEtran}
\IEEEoverridecommandlockouts

\usepackage{graphicx}
\usepackage{textcomp}
\usepackage{array}
\usepackage{epstopdf}
\usepackage{subfigure}
\usepackage{caption}
\usepackage{multirow}
\usepackage{titlesec}
\usepackage{setspace}
\usepackage{algorithm}
\usepackage{algorithmic}
\usepackage[misc]{ifsym}
\usepackage{amsmath}
\usepackage{listings}
\usepackage{xcolor}
\usepackage{framed} 

\usepackage{url}

\input{mymacros}

\begin{document}

\title{Boosting the Capability of Intelligent Vulnerability Detection by Training in a Human-Learning Manner}

\author{
\IEEEauthorblockN{
Shihan Dou\IEEEauthorrefmark{2}\IEEEauthorrefmark{6}, 
Yueming Wu\thanks{$\|$Co-first authors}\thanks{*Corresponding author}\IEEEauthorrefmark{3}\IEEEauthorrefmark{6}\IEEEauthorrefmark{1}, 
Wenxuan Li\IEEEauthorrefmark{2}, 
Feng Cheng\IEEEauthorrefmark{4}, 
Wei Yang\IEEEauthorrefmark{5}, 
Yang Liu\IEEEauthorrefmark{3}}
\IEEEauthorblockA{
\IEEEauthorrefmark{2}Fudan University,
\IEEEauthorrefmark{3}Nanyang Technological University\\
\IEEEauthorrefmark{4}Huazhong University of Science and Technology,
\IEEEauthorrefmark{5}University of Texas at Dallas}
}

\maketitle


\input{outline/abstract}
\input{outline/intro}

\input{outline/motivation}
\input{outline/system}

\input{outline/evaluation}

\input{outline/casestudy}

\input{outline/discussion}

\input{outline/relatedwork}
\input{outline/conclusion}

\bibliography{Humer}
\bibliographystyle{IEEEtran}
\end{document}

%% file: mymacros.tex
\newcommand{\ie}{\textit{i.e.,} }
\newcommand{\eg}{\textit{e.g.,} }

\newcommand\opt[1]{}
\newcommand\find[1]{}

\newcommand{\ls}[1]
   {\dimen0=\fontdimen6\the\font 
    \lineskip=#1\dimen0
    \advance\lineskip.5\fontdimen5\the\font
    \advance\lineskip-\dimen0
    \lineskiplimit=.9\lineskip
    \baselineskip=\lineskip
    \advance\baselineskip\dimen0
    \normallineskip\lineskip
    \normallineskiplimit\lineskiplimit
    \normalbaselineskip\baselineskip
    \ignorespaces
   }

\newenvironment{smalldescription}{
   \setlength{\topsep}{0pt}
   \setlength{\partopsep}{0pt}
   \setlength{\parskip}{0pt}
   \begin{description}
   \setlength{\leftmargin}{.2in}
   \setlength{\parsep}{0pt}
   \setlength{\parskip}{0pt}
   \setlength{\itemsep}{0pt}}{\end{description}}



%% file: outline/abstract.tex
\begin{abstract}

Due to its powerful automatic feature extraction, deep learning (DL) has been widely used in source code vulnerability detection.
However, although it performs well on artificial datasets, its performance is not satisfactory when detecting real-world vulnerabilities due to the high complexity of real-world samples.
Meanwhile, almost all DL-based methods consider only designing a new model to detect vulnerabilities and ignore the importance of the training process.
During their training phases, all training data are treated equally and fed into their models for training in a completely random order.
However, the understanding ability of a model is different when learning different vulnerability knowledge.
In other words, some vulnerability samples are difficult for the model to master while some samples are easy to learn.

In this paper, we propose to train DL-based vulnerability detection models in a human-learning manner, that is, start with the simplest samples and then gradually transition to difficult knowledge.
Specifically, we design a novel framework (\ie \emph{Humer}) that can enhance the detection ability of DL-based vulnerability detectors.
To validate the effectiveness of \emph{Humer}, we select five state-of-the-art DL-based vulnerability detection models (\ie \emph{TokenCNN}, \emph{VulDeePecker}, \emph{StatementGRU}, \emph{ASTGRU}, and \emph{Devign}) to complete our evaluations.
Through the results, we find that the use of \emph{Humer} can increase the F1 of these models by an average of 10.5\%.
Moreover, \emph{Humer} can make the model detect up to 16.7\% more real-world vulnerabilities.
Meanwhile, we also conduct a case study to uncover vulnerabilities from real-world open source products by using these enhanced DL-based vulnerability detectors.
Through the results, we finally discover 281 unreported vulnerabilities in NVD, of which 98 have been ``silently” patched by vendors in the latest version of corresponding products, but 159 still exist in the products.
\end{abstract}


%% file: outline/intro.tex
\section{Introduction}

Recently, as the complexity of computer software systems has increased, the number of potential security vulnerabilities has shown an increasing trend.
In 2021, \emph{National Vulnerability Database} (NVD) reported that, starting from 2018, the number of disclosed vulnerabilities has exceeded 15,000 for four consecutive years \cite{nvd-statistical}.
Although some of these vulnerabilities have been fixed, security incidents caused by vulnerabilities are still emerging in endlessly.
For example, in May 2017, ``WannaCry" ransomware \cite{wannacry} exploited a dangerous vulnerability (\ie ``EternalBlue") disclosed by \emph{National Security Agency} (NSA) in March of the same year to attack Windows operating system and invade users' hosts.
The ransomware has spread to at least 150 countries and regions, causing hundreds of millions of losses to governments, enterprises, universities, and other industries.
To resist these security threats, source code vulnerability detection has become the most effective solution.

Generally speaking, source code vulnerability detection methods can be classified into two main categories (\ie the code-similarity-based methods \cite{jang2012redebug, kim2017vuddy, pham2010detection, li2012cbcd, li2016vulpecker}
and the pattern-based methods \cite{checkmarx, rats, flawfinder}).
Methods in the first category focus on detecting vulnerabilities caused by code cloning.
They leverage different similarity calculation techniques to measure the similarity between known vulnerabilities and the source code to be detected.
If the similarity is higher than a preset threshold (\eg 80\%), the source code will be reported as a vulnerability.
In other words, code-similarity-based methods can only detect known vulnerabilities. 

To detect more vulnerabilities, pattern-based methods are designed.
Among these methods, deep learning (DL)-based approaches \cite{2018VulDeePecker, zou2019muvuldeepecker, zhou2019devign, lin2017poster, duan2019vulsniper, li2021sysevr, russell2018automated, wang2020funded} have attracted wide attention since they do not require experts to manually define features and can automatically generate vulnerability patterns from program source code. 
For example, \emph{VulDeePecker} \cite{2018VulDeePecker} treats the program slices of source code as text and applies a \emph{bidirectional long short-term memory} (BLSTM) neural network to detect vulnerabilities.
\emph{Funded} \cite{wang2020funded} first conducts complex program analysis to extract augmented \emph{abstract syntax tree} (AST) of source code, and then uses a \emph{graph neural network} (GNN) to train a vulnerability detector.
According to their papers' results, they can achieve ideal performance on detecting vulnerabilities in \emph{Software Assurance Reference Dataset} (SARD) \cite{sard}.
But in fact, vulnerabilities in SARD are synthetic and are only used for simple academic research.
To check the feasibility of DL-based methods on real-world vulnerability detection, a recent study \cite{chakraborty2021deep} uses several real-world vulnerability datasets and conducts a comparative evaluation on certain DL-based vulnerability detectors.
Through the results, they find that due to the high complexity of real-world vulnerabilities, the detection performance of evaluated detectors can drop a lot.

Meanwhile, we also observe that almost all DL-based vulnerability detectors only focus on designing a new model and neglect the importance of their training phases.
They train models in a straightforward manner, that is, all training samples are treated equally and presented in a completely random order during training.
However, a model has different understanding capabilities when learning different vulnerability knowledge.
It can converge quickly when learning a certain type of vulnerability knowledge, but it may converge slowly when targeting some other types of vulnerability knowledge (Section II).
In other words, some vulnerability samples are difficult for the model to understand and some samples are easy to learn.
For humans, there is a similar phenomenon in our learning process, that is, our ability to understand different knowledge is also different.
However, in our human learning process, we do not directly learn difficult knowledge at the beginning, but start from understanding simple knowledge, and then transition to the difficult content.
That is to say, our human learning method is phased, unlike DL-based vulnerability detection, where samples are completely randomly inputted into the model for training.
In fact, previous studies \cite{skinner1958reinforcementCL1, elman1993learningCL2, peterson2004dayCL3, krueger2009flexibleCL4} have shown that if a machine learning model can be trained in a phased learning way, its learning effect will be better.
Researchers refer to this type of learning method as \emph{Curriculum Learning}, which is first proposed by Bengio \emph{et al.} \cite{bengio2009curriculum}.
Its original formulation can be viewed as a continuation method, which can help to find better local minima of a highly non-convex criterion and can reach faster convergence to a minimum of the training criterion \cite{bengio2009curriculum}.

In this paper, we draw inspiration from \emph{Curriculum Learning} and propose to model human learning procedures to train DL-based vulnerability detectors.
Specifically, we not only simulate human phased learning, but also consider the review and consolidation in human learning process to train DL-based vulnerability detectors.
For knowledge phased learning, we introduce two difficulty computation techniques of vulnerability samples.
The first is to start from the model itself and focus on the model's ability to recognize vulnerability samples.
The second is to start from the vulnerability sample and focus on the code complexity of the sample itself.
After sorting all samples by difficulty, we divide them into different buckets according to the degree of difficulty, and then start training from the simplest bucket.
For knowledge review, before learning the next bucket, we put the samples of the previous bucket into this bucket first, and then learn with this bucket to achieve the purpose of review.
For knowledge consolidation, we record all mispredicted samples in the training process of each bucket, and apply semantic-equivalent code transformations to generate multiple variants of these samples.
These variant samples constitute an \emph{error book} and are used to fine-tune the model to strengthen the learning of ununderstood knowledge.

We implement a novel framework namely \emph{Humer} and select five state-of-the-art DL-based vulnerability detectors (\ie \emph{TokenCNN} \cite{russell2018automated}, \emph{VulDeePecker} \cite{2018VulDeePecker}, \emph{StatementGRU} \cite{lin2019statementgru}, \emph{ASTGRU} \cite{feng2020astgru}, and \emph{Devign} \cite{zhou2019devign}) to evaluate \emph{Humer}.
Through the results, we find that using \emph{Humer} to train these models can make them detect 12.6\% more vulnerabilities on average.
Meanwhile, the F1 values also increase by an average of 10.5\% after using \emph{Humer}.
To further demonstrate the effectiveness of \emph{Humer}, we conduct a case study on seven popular open source products with a total number of lines of code exceeding 58 million.
Specifically, we first train five models (\ie \emph{TokenCNN} \cite{russell2018automated}, \emph{VulDeePecker} \cite{2018VulDeePecker}, \emph{StatementGRU} \cite{lin2019statementgru}, \emph{ASTGRU} \cite{feng2020astgru}, and \emph{Devign} \cite{zhou2019devign}) with \emph{Humer} to obtain enhanced detectors, and then scan vulnerabilities on downloaded products.
Through the scanning reports, we finally detect 281 vulnerabilities that are not reported in NVD. 
Among them, 98 have been ``silently'' patched by vendors in the latest version of corresponding products, 24 vulnerabilities have been deleted, and the other 159 still exist in the products.
As for improvement, the scanning results indicate that using \emph{Humer} can make them (\ie \emph{TokenCNN}, \emph{VulDeePecker}, \emph{StatementGRU}, \emph{ASTGRU}, and \emph{Devign}) detect 37 more unreported vulnerabilities on average.
 
\par In summary, this paper makes the following contributions:
\begin{itemize}
\item{As far as we know, we are the first to model human learning procedures into the training of DL-based vulnerability detection.}
\item{We implement \emph{Humer}, a model-agnostic framework that can enhance the effectiveness of DL-based vulnerability detection models.}
\item{We select five state-of-the-art DL-based vulnerability detectors (\ie \emph{TokenCNN}, \emph{VulDeePecker}, \emph{StatementGRU}, \emph{ASTGRU}, and \emph{Devign}) and conduct evaluations using a real-world vulnerability dataset.
Experimental results report that the use of \emph{Humer} can make them detect 12.6\% more vulnerabilities on average.}
\item{We conduct a case study on more than 58 million lines of code to further demonstrate the practicability of \emph{Humer}.
Through the scanning results, we discover 281 vulnerabilities that are not reported in NVD.}
\end{itemize}


%% file: outline/motivation.tex
\section{Motivation}

Due to the powerful capability of automatic feature extraction, deep learning (DL) has been widely used in vulnerability detection.
However, in these DL-based vulnerability detectors, all training samples are randomly presented to the model, ignoring the complexity of the vulnerability samples and the current learning state of the model.
In fact, we humans do not learn like this, but start learning from simple knowledge, and then move on to difficult knowledge.
The main reason why we humans learn this way is because our ability to understand different knowledge is different. 
Some knowledge can be easily mastered, while some knowledge we need to spend more time to understand.
For those difficult knowledge, we should first learn their basic principles (simple knowledge). 
After mastering these principles, we can understand the difficult knowledge faster and more thoroughly.
In order to check whether there are similar phenomena in DL-based vulnerability detection models, we perform a simple investigation in this section.
Specifically, we mainly answer the following questions:

\emph{Is the ability of DL-based vulnerability detection models to learn different vulnerability knowledge different?}

\begin{figure*}
\centering
\subfigure{
\begin{minipage}[t]{0.3\textwidth}
\centering
\includegraphics[width=\textwidth]{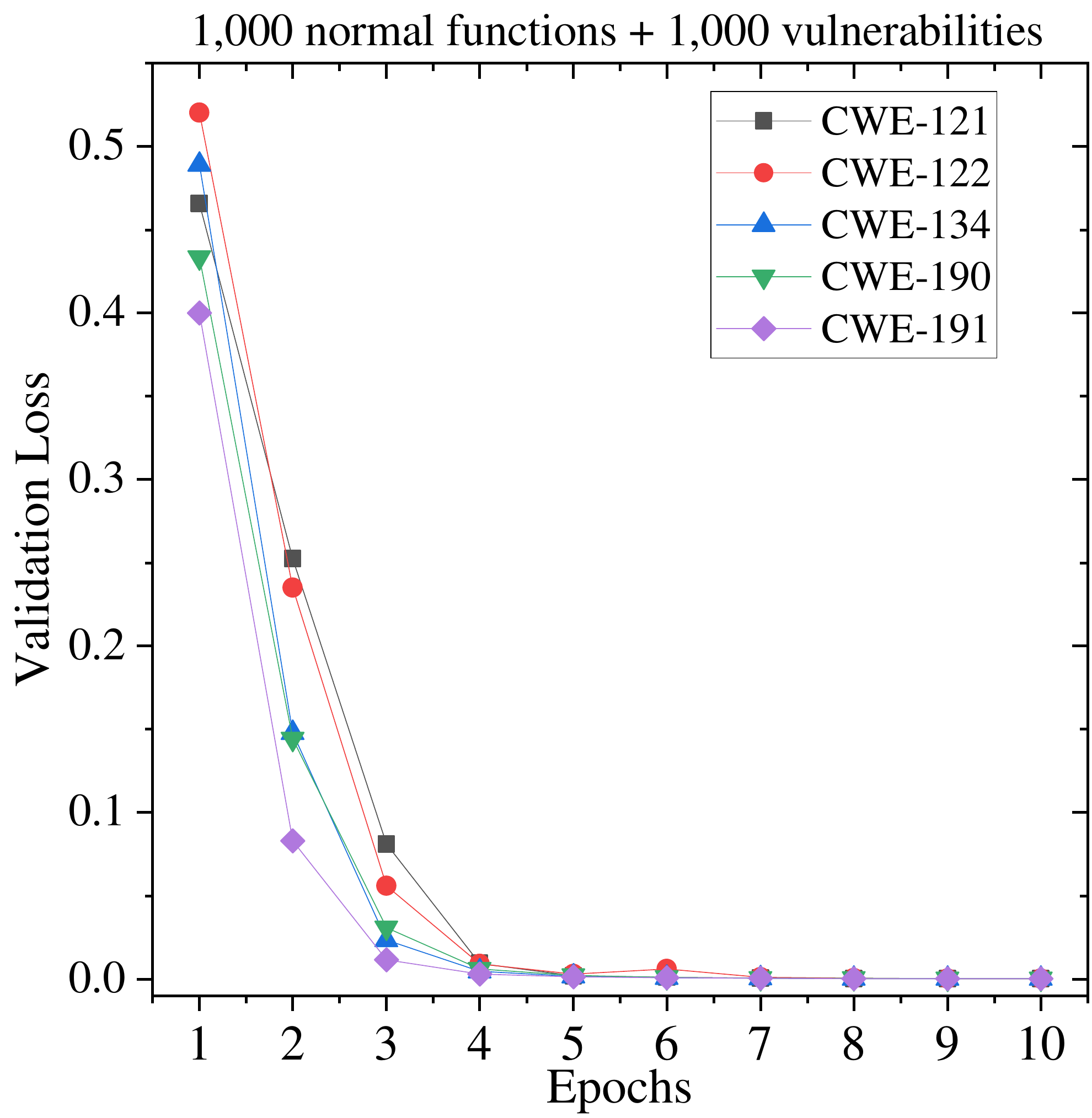}
\end{minipage}
}
\subfigure{
\begin{minipage}[t]{0.3\textwidth}
\centering
\includegraphics[width=\textwidth]{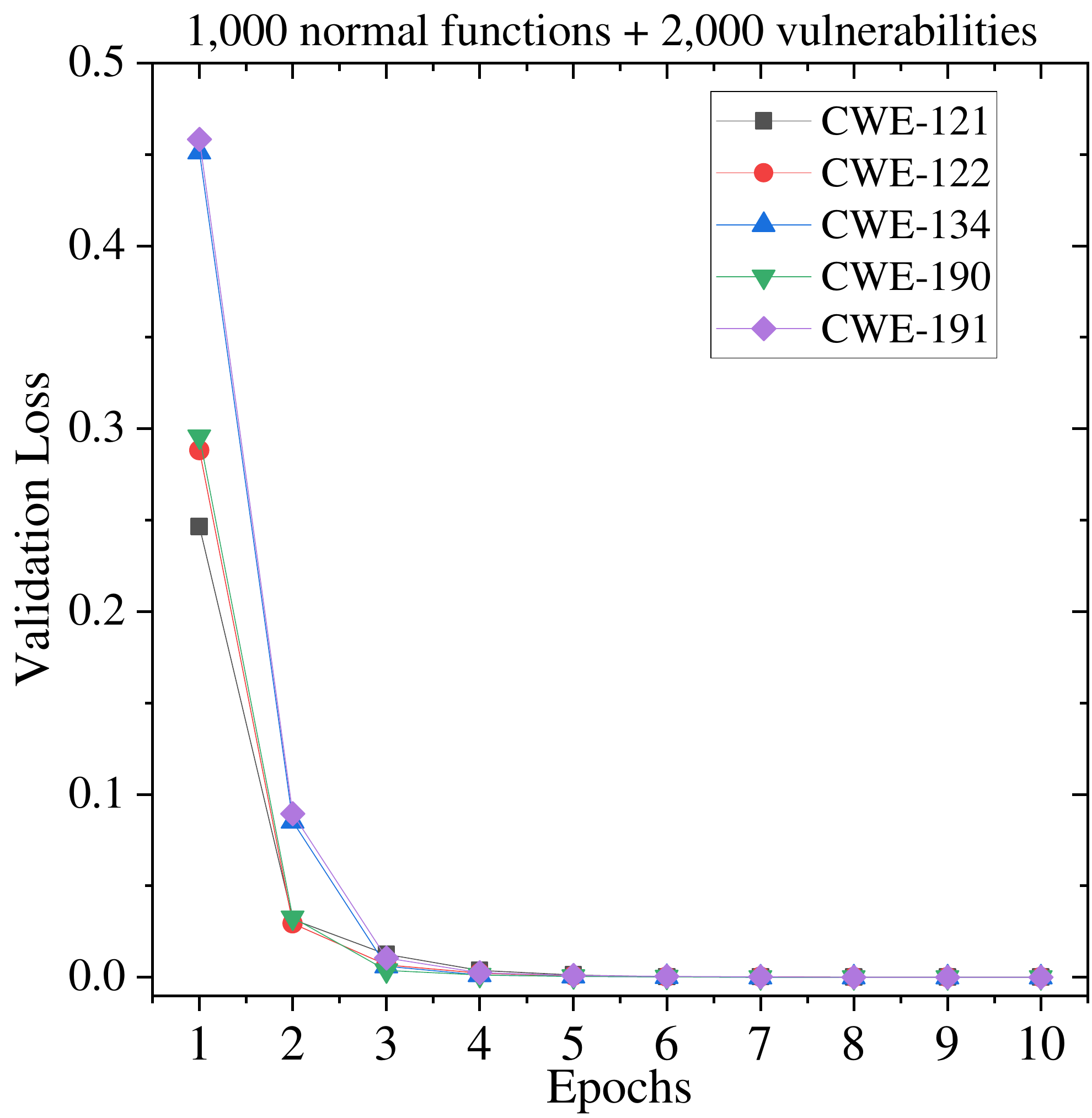}
\end{minipage}
}
\subfigure{
\begin{minipage}[t]{0.3\textwidth}
\centering
\includegraphics[width=\textwidth]{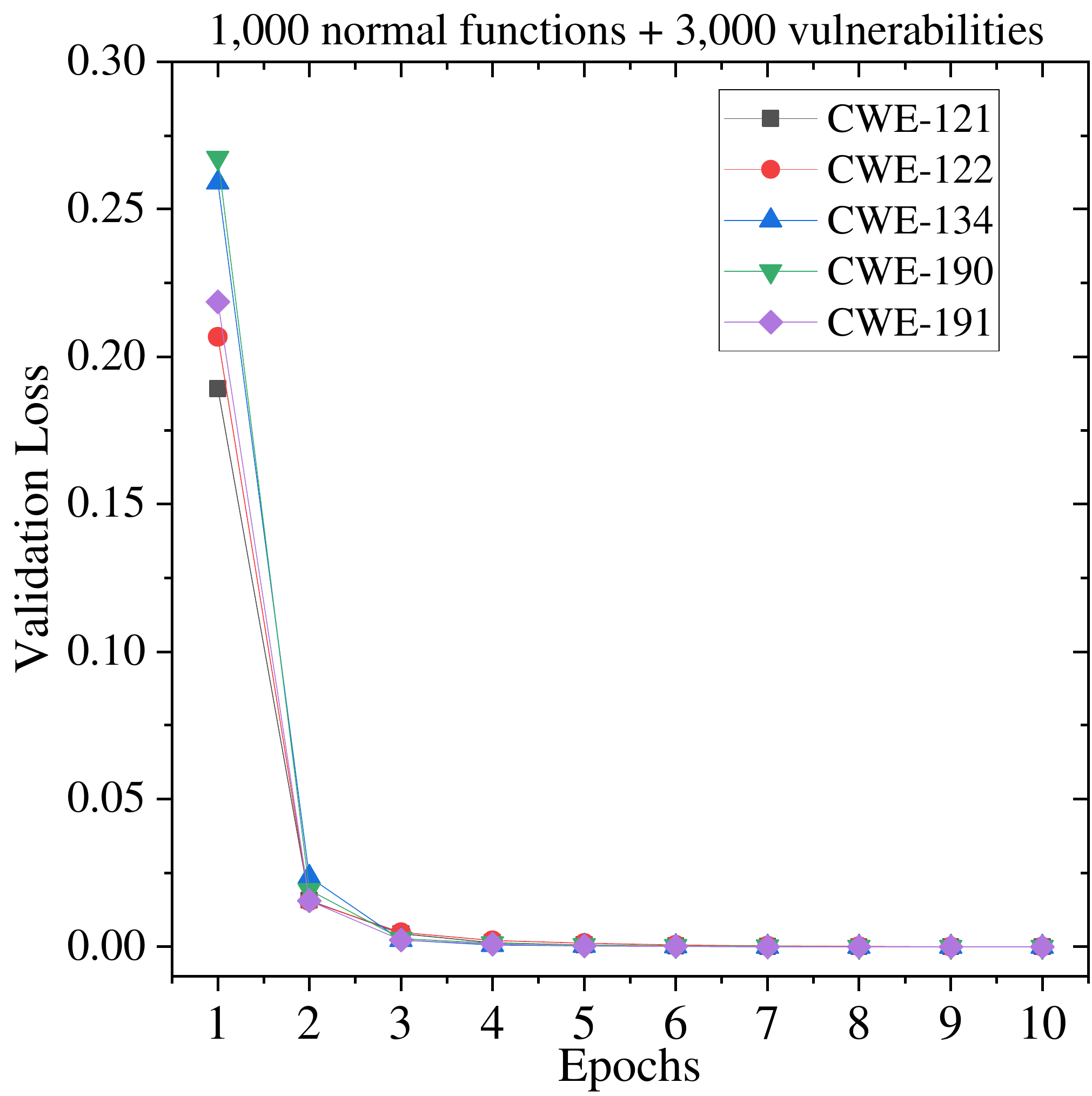}
\end{minipage}
}

\caption{Validation losses of \emph{TokenCNN} on detecting different types of vulnerabilities}
\label{fig:cwe}
\end{figure*}

To answer the raised question, we use the vulnerability data in a prior study \cite{lin2019statementgru} as the test object.
The authors collect a total of 74 different types of vulnerable functions from SARD, and we select the vulnerability types with more than 3,100 vulnerabilities.
There are five types of vulnerabilities that meet our criteria (\ie CWE-121, CWE-122, CWE-134, CWE-190, and CWE-191).
The descriptions of these five types of vulnerability are described in Table \ref{tab:cwe}.
To commence our research, we randomly choose 3,100 vulnerability samples from each type of vulnerability data, and randomly select 1,100 samples from normal data to form our experimental dataset.
For each type of vulnerability, we randomly select 100 normal functions and 100 vulnerabilities as the validation set, and other samples (1,000 normal functions and 3,000 vulnerabilities) as the training set.
We mainly conduct three experiments. 
The first is to train a model by using 1,000 normal functions and 1,000 vulnerabilities and then record the loss during validation phase.
The second is to add 1,000 vulnerabilities to the training set and then use these samples to train the model and record the loss during validation phase.
The third is to continue to add 1,000 vulnerabilities to the training set and then repeat the previous process.
For DL-based vulnerability detection model, we choose \emph{TokenCNN} \cite{russell2018automated} since it has been widely used as a comparative model by many other researchers \cite{zhou2019devign, chakraborty2021deep}.
It first applies lexical analysis to parse the source code of a program into tokens and then inputs them into a convolutional neural network to train a vulnerability detector.

\begin{table}[htbp]
  \centering
  \small
  \caption{The descriptions of selected five vulnerability types}
    \begin{tabular}{|m{1.3cm}<{\centering}|m{6.6cm}<{\centering}|}
    \hline
    CWE   & Descriptions \\
    \hline
    CWE-121 & A stack-based buffer overflow condition is a condition where the buffer being overwritten is allocated on the stack. \\
    \hline
    CWE-122 & A heap overflow condition is a buffer overflow, where the buffer that can be overwritten is allocated in the heap portion of memory. \\
    \hline
    CWE-134 & The software uses a function that accepts a format string as an argument, but the format string originates from an external source. \\
    \hline
    CWE-190 & The software performs a calculation that can produce an integer overflow or wraparound, when the logic assumes that the resulting value will always be larger than the original value. \\
    \hline
    CWE-191 & The product subtracts one value from another, such that the result is less than the minimum allowable integer value, which produces a value that is not equal to the correct result. \\
    \hline
    \end{tabular}%
  \label{tab:cwe}%
\end{table}%

The experimental results are shown in Figure \ref{fig:cwe}. 
It can be seen that although the size of the training set is the same, the model has different learning capabilities for different types of vulnerability (\ie the loss is different).
When the vulnerability data is added, the loss generally becomes lower, indicating that the model learns more vulnerability knowledge.
But for some types of vulnerabilities, even if the number of vulnerabilities increases to 2,000, the model's ability is still not good enough.
For example, for CWE-191, when the training set consists of 2,000 vulnerabilities, the loss in the first epoch is 0.4583.
For CWE-190, even if there are only 1,000 vulnerabilities in the training set, its loss in the first epoch is lower than 0.4583, only 0.4332.
Such results indicate that \emph{TokenCNN} has different understanding capabilities when learning different types of vulnerability knowledge.
Compared to CWE-191, \emph{TokenCNN} has a stronger understanding ability on learning knowledge in CWE-190.
Based on the observation, we propose to train DL-based vulnerability detection models in a meaningful order rather than a random order.

%% file: outline/system.tex
\section{Design of Humer}
In this section, we introduce \emph{Humer}, a novel framework that models human learning procedures to enhance the effectiveness of DL-based vulnerability detection.

\begin{figure*}[htbp]
\centerline{\includegraphics[width=0.9\textwidth]{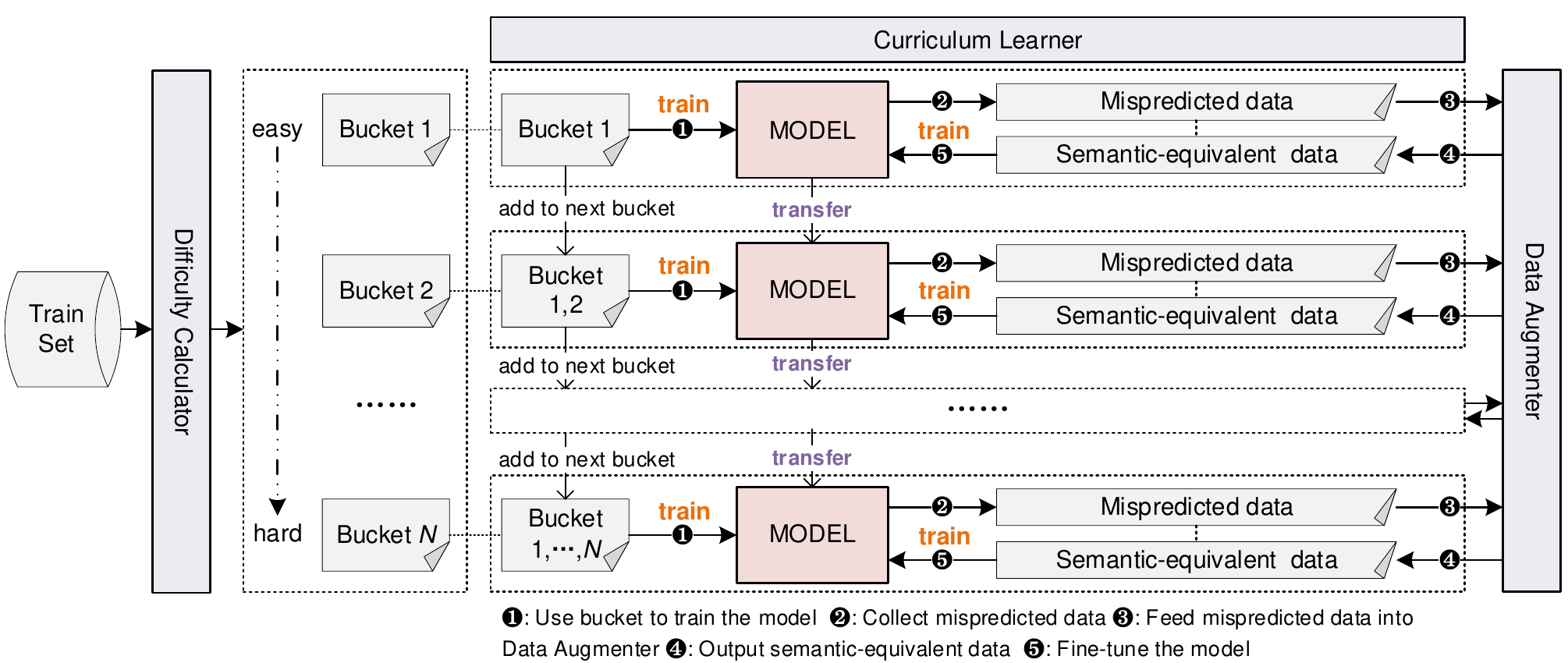}}
\caption{\small{Framework overview of \emph{Humer}}}
\label{fig:overview}
\end{figure*}

\subsection{Overview}
\par As shown in Figure \ref{fig:overview}, \emph{Humer} consists of three main components: \emph{Difficulty Calculator}, \emph{Curriculum Learner}, and \emph{Data Augmenter}.
\begin{itemize}
\item \emph{Difficulty Calculator:}
Given a training dataset, we first compute the difficulty of each sample and divide them into several buckets according to the degree of difficulty.
\item \emph{Curriculum Learner:}
Given sorted buckets, we start training from the simplest bucket, and then learn the more difficult buckets in turn.
For each bucket, we record all mispredicted data during training and use \emph{Data Augmenter} to generate the corresponding variants, and then fine-tune the model using these variants.
\item \emph{Data Augmenter:}
Given mispredicted data, we apply semantic-equivalent code transformations to construct the corresponding variant data.
\end{itemize}

\subsection{Difficulty Calculator}

Because a model has different abilities on understanding different vulnerability knowledge, we compute a sample's difficulty by analyzing the model's ability on recognizing it.
However, the runtime overhead of analyzing a model is not lightweight.
Therefore, we design another alternative model-independent difficulty calculation method from the perspective of we humans.
That is, we compute the difficulty of a sample by analyzing the source code complexity directly.

\subsubsection{Model-based Difficulty Calculator}
In this part, we pay attention to the model's ability on recognizing vulnerability samples.
In other words, a sample with a stronger recognition ability will be considered easy to learn, while a sample that is not easily recognized will be considered a difficult sample.

\begin{figure}[htbp]
\centerline{\includegraphics[width=0.48\textwidth]{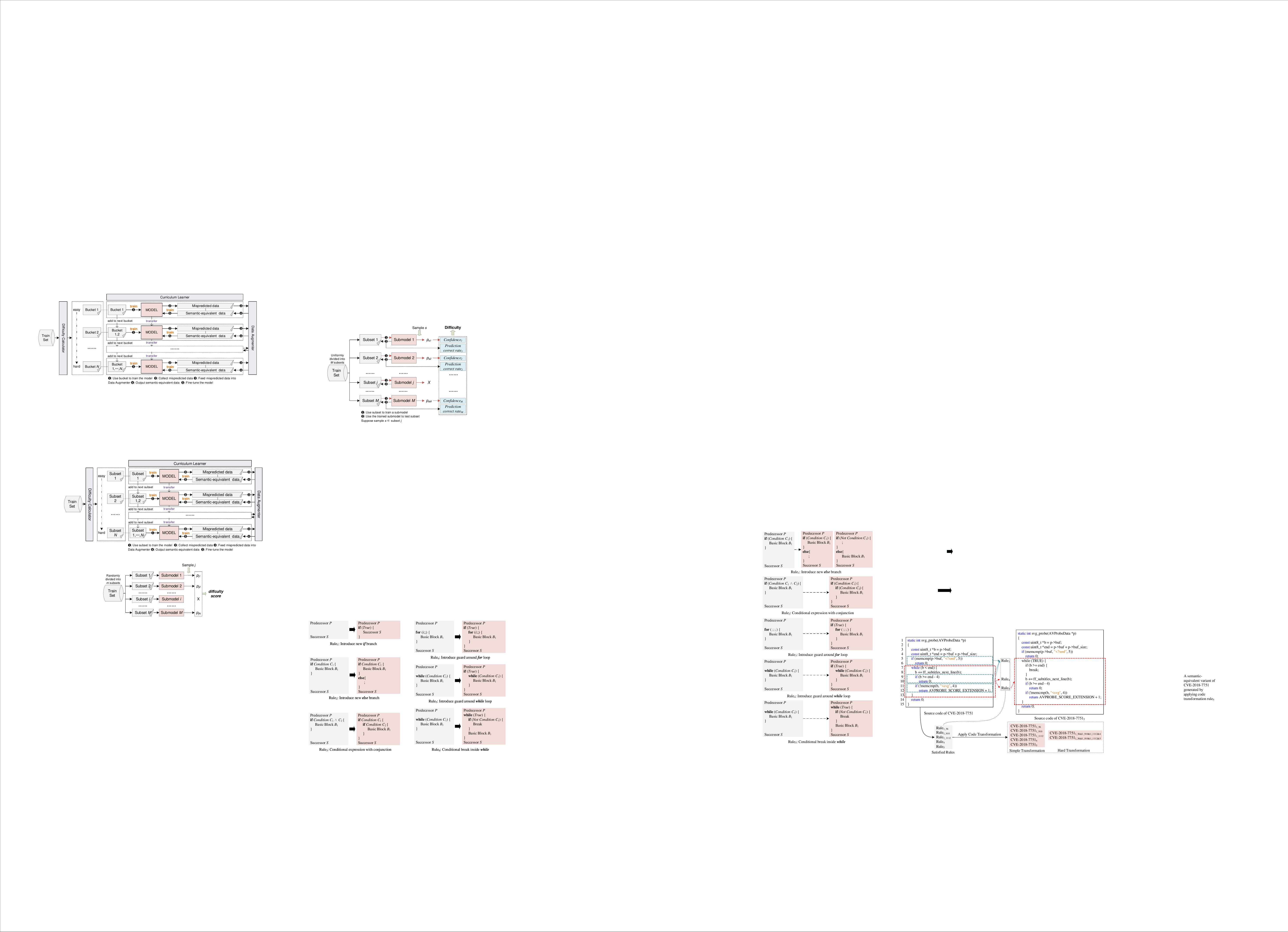}}
\caption{Model-based difficulty calculator: the training set is split into $M$ subsets, after the submodels are trained on them, each sample will be predicted by all other submodels (except the one it belongs to).}
\label{fig:model-based}
\end{figure}

To complete difficulty calculation, we first split our training set $D$ into $M$ subsets uniformly as $\left \{ d_{i}:i=1,2,...,M \right \}$, and then train $M$ corresponding submodels $\left \{ m_{i}:i=1,2,...,M \right \}$.
Note that each submodel only sees 1/$M$ samples in the training set.
After obtaining $M$ submodels, we input samples in $D$ to these submodels to obtain the corresponding predicted probabilities.
The model considers a sample as a vulnerability only when the predicted probability is greater than 0.5. 
Therefore, we take the part with the predicted probability greater than 0.5 as the submodel’s \emph{confidence} of identifying the vulnerability sample.
If it is a vulnerability sample, the greater the predicted probability, the greater the \emph{confidence} that the submodel classifies the sample as a vulnerability.
On the contrary, if the sample is normal, the lower the predicted probability, the greater the \emph{confidence} that the submodel distinguishes the sample as normal.

In addition, for each submodel, its prediction \emph{correct rate} generally does not reach 100\%, that is, some samples may be predicted incorrectly.
For example, assuming that subset $d_{1}$ consists of 1,000 vulnerability samples and 1,000 normal samples.
After completing the training of submodel $m_{1}$ by using these samples, we then input them to submodel $m_{1}$ for testing.
In other words, these 2,000 samples are both training set and testing set.
For the testing results, suppose $m_{1}$ predicts that 1,100 samples are vulnerabilities, but in fact, only 800 samples are vulnerabilities, and the other 300 samples are normal samples.
It means that the \emph{correct rate} of submodel $m_{1}$ in predicting vulnerabilities is 800/1100=72.7\%, and there is a 27.3\% probability that the prediction is wrong.

To obtain more credible computation results, we consider both the prediction \emph{correct rate} of the submodel and the discrimination \emph{confidence} of the sample.
That is, the higher the prediction \emph{correct rate} of the submodel itself and the greater the \emph{confidence} in distinguishing the sample, the stronger the model's ability to recognize the sample.
The calculation formula is as follows:

\begin{equation}
\label{model_difficulty}
DS_{s\in d_{j}}=\left\{
\begin{aligned}
-\sum_{i\neq j}^{N}\frac{TP_i}{TP_i+FP_i}*(p_{i}-0.5), && label(s) = 1\\
-\sum_{i\neq j}^{N}\frac{TN_i}{TN_i+FN_i}*(0.5-p_{i}), && label(s) = 0
\nonumber
\end{aligned}
\right.
\end{equation}

For each subset $d_i$, we first use it to train a submodel $m_i$.
After training, we then input it to $m_i$ for testing and record the \emph{true positives} (\ie TP, \#samples correctly predicted as vulnerable), \emph{true negatives} (\ie TN, \#samples correctly predicted as normal), \emph{false positives} (\ie FP, \#samples incorrectly predicted as vulnerable), and \emph{false negatives} (\ie FN, \#samples incorrectly predicted as normal).
By this, we can obtain the prediction \emph{correct rate} of all submodels.
More specifically, for each submodel, we compute the \emph{correct rate} in predicting vulnerabilities by analyzing the quotient of TP and (TP+FP).
For normal samples, we compute the quotient of TN and (TN+FN) as the prediction \emph{correct rate}.
As shown in Figure \ref{fig:model-based}, given a sample $s$, suppose it belongs to subset $d_j$ and its label is 1 (\ie $s$ is a vulnerability).
To compute its difficulty, we input it to all submodels except submodel $m_j$ because $m_j$ has already seen it during training.
For each submodel, after obtaining the corresponding predicted probability $p$ of $s$, we can calculate the discrimination \emph{confidence} by computing $p$ minus 0.5 since $s$ is a vulnerability sample.
After finishing computing the prediction \emph{correct rate} of the submodel and the discrimination \emph{confidence} of the sample, the submodel's ability on recognizing the sample will be computed as their product.  
Each submodel will calculate a recognition ability score, and the negative of the sum of these scores (\ie $N$-1 scores) is the difficulty score of the sample.
The lower the difficulty score, the stronger the model's ability to distinguish the sample, making it easier to learn.

\subsubsection{Code-based Difficulty Calculator}
In this part, we compute the difficulty of a sample by analyzing the code complexity from the source code directly.
Samples with lower code complexity are regarded as simple samples, and samples with higher complexity are treated as difficult samples.

According to prior work \cite{coleman1994mi01, oman1992mi02}, the maintainability of a program can give an overall picture of program complexity.
The lower complexity of the program, the easier it is to maintain.
To complete our difficulty calculation, we adopt \emph{Maintainability Index} as the standard to evaluate the code complexity of samples.
\emph{Maintainability Index} \cite{heitlager2007mi03, coleman1995mi04, welker1997mi05} is more of an empirical measure, it has been developed over the years with consultants from Hewlett-Packard and its software teams \cite{codecomplexity}. 
This index weighs \emph{Cyclomatic Complexity} \cite{mccabe1976complexity} and \emph{Halstead Volume} \cite{halstead1977elements} against the number of \emph{Source Lines of Code} in the program. 
These three metrics are described as follows:

\begin{itemize}
\item \emph{Source Lines of Code}:  
It is the most straightforward metric used to measure the size of the program.
It is obtained by counting the total number of lines in the source code excluding comments and blank lines.
The more lines, the more complex the code.
\item \emph{Cyclomatic Complexity}:
It measures how much control flow exists in a program. 
It is a count of the number of decisions in the source code.
The higher the count, the more complex the code. 
\item \emph{Halstead Volume}:
It measures how much information is in the source code. 
It looks at how many variables are used and how often they are used in the source code.
The larger the volume, the more complex the code.
\end{itemize}

According to prior research \cite{heitlager2007mi03, coleman1995mi04, welker1997mi05}, if we define the \emph{Source Lines of Code} as $L$, the \emph{Cyclomatic Complexity} as $G$, and the \emph{Halstead Volume} as $V$, then the \emph{Maintainability Index} (MI) can be calculated as follows:

\begin{equation}
\label{model_difficulty}
MI=171-5.2\ln V-0.23G-16.2\ln L
\nonumber
\end{equation}

The more complex the code, the lower the \emph{Maintainability Index}, and the greater the difficulty score.

\subsection{Curriculum Learner}
After evaluating the difficulty of all samples in the training set, we sort them by their difficulty scores and divide them into $N$ buckets uniformly as $\left \{ b_{i}:i=1,2,...,N \right \}$.
By this, samples in the training set are collected into $N$ different levels of difficulty, ranging from $b_{1}$ (the easiest) to $b_{N}$ (the hardest).

After completing the bucket assignment, we start our training from the easiest bucket. 
When the training reaches convergence or a preset number of a maximum epoch, we stop it and transfer to the next training stage.
Specifically, the preset maximum number of epochs in our experiments is 10, and the training will be treated as converged if the training loss stops to decrease.
Before starting the next training, we first add the samples of the previous bucket to the next bucket and apply transfer learning to learn this bucket, so as to realize the knowledge transfer between the two buckets.
Since we allocate $N$ buckets, we perform a total of $N$-1 transfer learning.

In our human learning process, it is difficult to understand all knowledge at once.
For the knowledge that we don't understand, we humans generally need to learn many times to master it.
We model this learning procedure into the training of DL-based vulnerability detection.
Specifically, for each bucket, we record the mispredicted samples during training, and leverage our \emph{Data Augmenter} module to generate the corresponding semantic-equivalent variant samples.
These variant samples constitute an \emph{error book} and are used to fine-tune the model to strengthen the learning of ununderstood knowledge.
The output of \emph{Curriculum Learner} is a model that can be used to detect vulnerability.

\subsection{Data Augmenter}
After collecting mispredicted samples, we feed them into our \emph{Data Augmenter} to output some semantic-equivalent variant samples.
The implementation of \emph{Data Augmenter} is based on several designed rules.
By using these rules, we can mutate the source code of mispredicted samples to generate new semantically equivalent samples.

\begin{figure}[htbp]
\centerline{\includegraphics[width=0.46\textwidth]{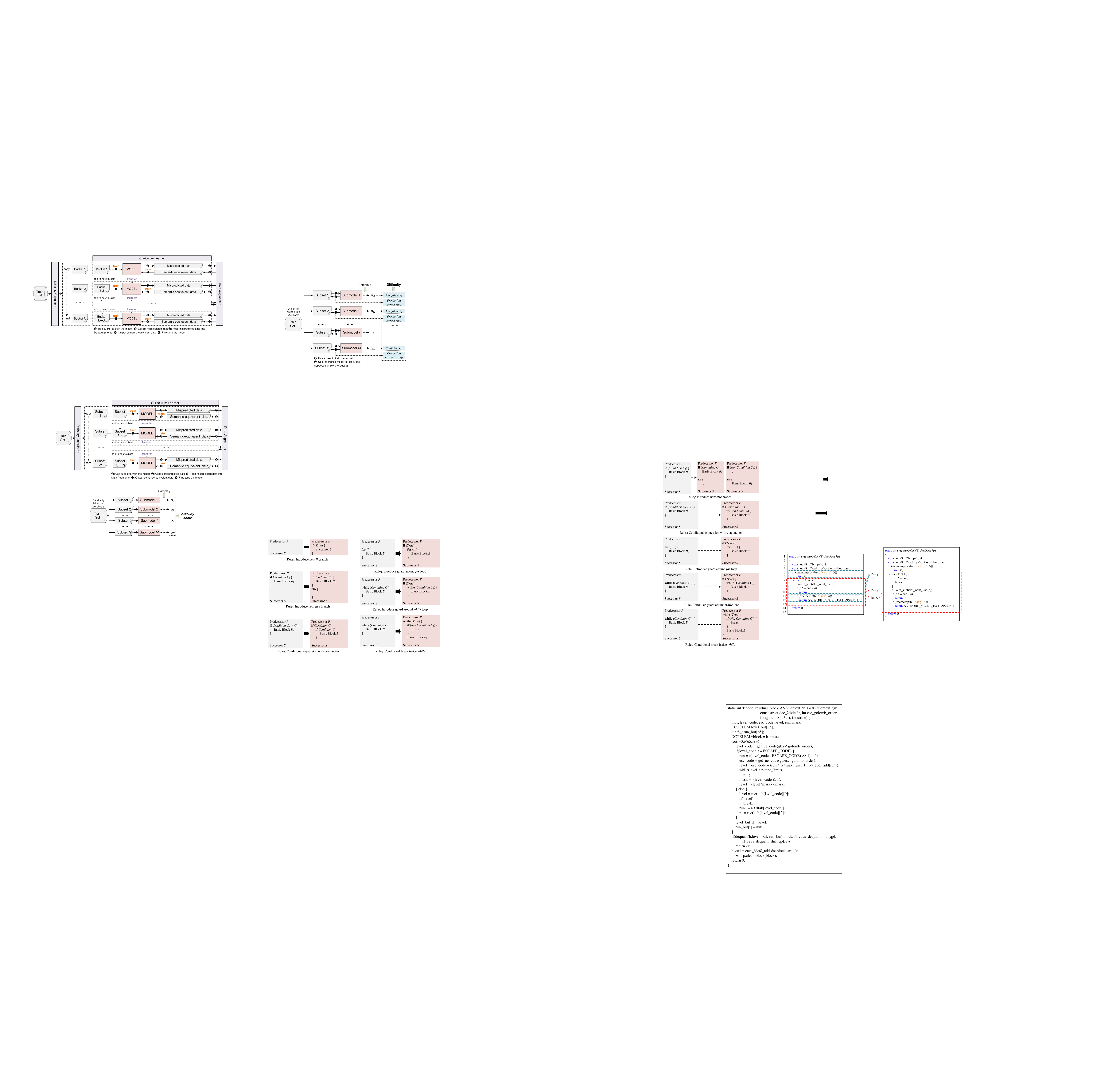}}
\caption{Code transformation rules used in \emph{Data Augmenter}}
\label{fig:rules}
\end{figure}

\begin{figure*}[htbp]
\centerline{\includegraphics[width=0.86\textwidth]{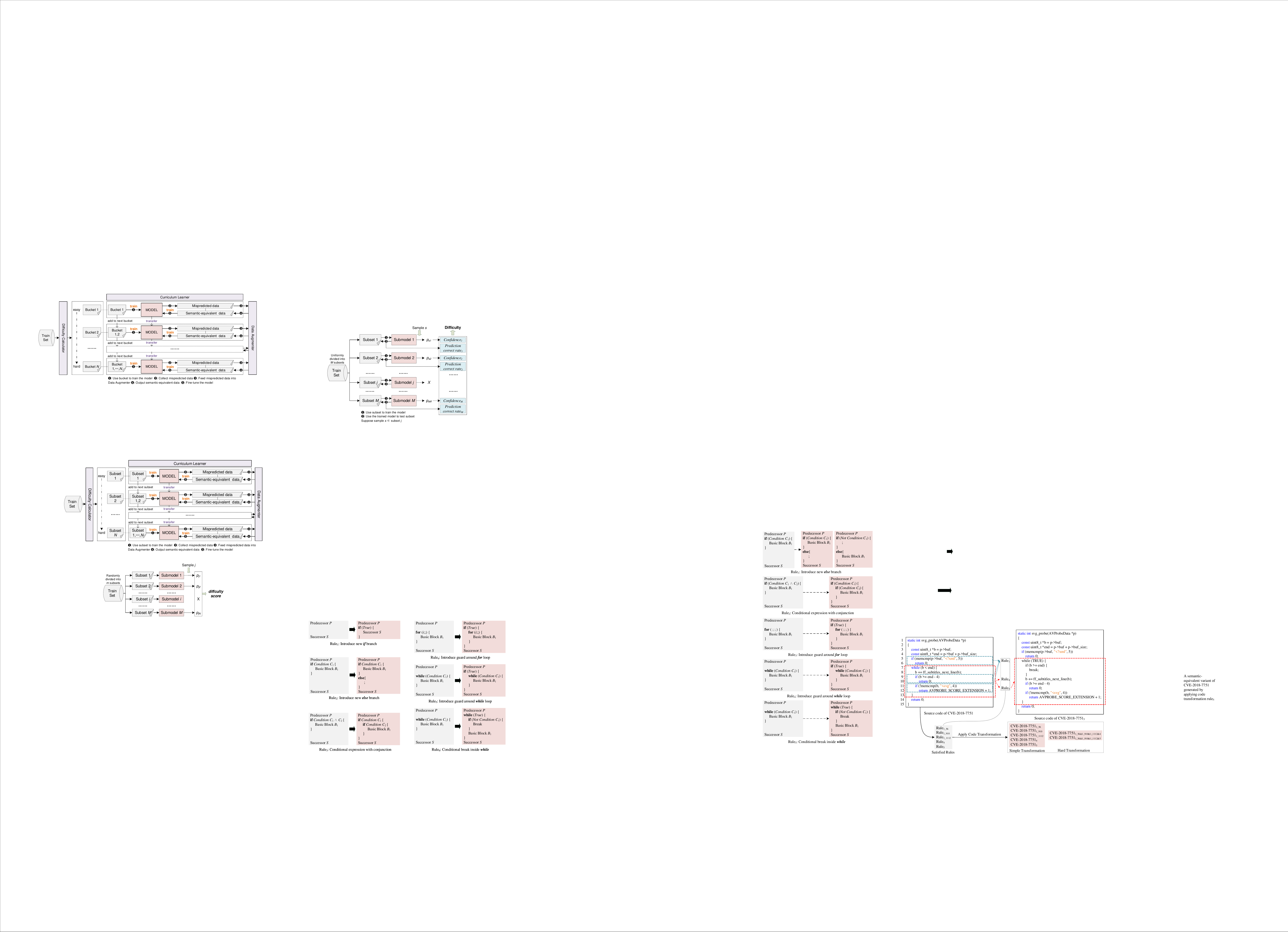}}
\caption{Mutating CVE-2018-7751 by applying code transformation rules}
\label{fig:rule5-example}
\end{figure*}

On the one hand, to make a model resilient to common code modifications while preserving program semantics, some DL-based vulnerability detection techniques apply abstraction and normalization to the source code.
For example, they may map user-defined variables to symbolic names in a one-to-one manner (\eg VAR).
On the other hand, to make a model more effective on vulnerability detection, researchers may consider distilling the program semantics into different code representations such as \emph{abstract syntactic tree} (AST) and \emph{control flow graph} (CFG).
These code representations will be fed into different neural networks such as \emph{graph neural network} (GNN) to train vulnerability detectors.
Therefore, we do not select those simple code transformations such as variable rename but design several rules that can generate new semantic-equivalent variants with different control flow structures.
Our rule design is based on the observation that the same algorithm can have syntactically different implementations.
More specifically, we implement a total of five code transformation rules as shown in Figure \ref{fig:rules}:
\begin{itemize}
\item \emph{$Rule_1$ (Introduce new else branch)}:
This rule introduces new $else$ branching statement to an existing $if$ conditional statement or reverse the condition in the $if$ conditional statement.
\item \emph{$Rule_2$ (Conditional expression with conjunction)}:
A conjunct $if$ statement with conditions $C_1$ and $C_2$ can be rewritten as a nested $if$ structure, which contains conditions $C_1$ and $C_2$ individually.
\item \emph{$Rule_3$ (Introduce guard around for loop)}:
A program containing a $for$ loop statement can be changed to a new program structure by introducing guard around the loop.
\item \emph{$Rule_4$ (Introduce guard around while loop)}:
Similar to $Rule_3$, a program containing a $while$ loop statement can be mutated into a new program structure by introducing guard around the loop.
\item \emph{$Rule_5$ (Conditional break inside while)}:
A program that loops until a condition $C_1$ is satisfied can be transformed into another infinite loop program, with a $C_1$ conditional break instruction inside the loop's body.
\end{itemize}

Given a program, we first traverse it to search for the statements that satisfy the code transformation rules.
After collecting all the code statements that can be mutated, we perform code transformation on these codes according to the conforming rules.
To implement code transformation with different levels of difficulty, we design two types of transformation combinations.
The first type is simple transformation, that is, transform only one code segment in the program that can be mutated.
The second type is hard transformation, that is, all codes that satisfy the transformation are transformed.

To better illustrate our code transformation phase, we take CVE-2018-7751 as an example and show the detailed procedure in Figure \ref{fig:rule5-example}.
After analyzing the program, we find that there are four code fragments in CVE-2018-7751 that meet the transformation conditions.
Specifically, lines 5 to 6, lines 9 to 10, and lines 11 to 12 all meet $Rule_1$, and lines 7 to 13 meet $Rule_4$ and $Rule_5$.
To obtain variant samples, we first use $Rule_1$ to apply code transformation on lines 5 to 6, and the other codes remain unchanged. 
In this way, a variant can be generated.
Repeating the above processes, we can construct five variants, which are called simple transformations.
To generate more difficult transformation codes, we transform all the codes that satisfy the transformation rules, which are called hard transformations.
In one word, after inputting CVE-2018-7751 into \emph{Data Augmenter}, we can obtain seven semantic-equivalent variant samples with different control flow structures.

%% file: outline/evaluation.tex
\section{Experiments}

Since \emph{Humer} is composed of three main modules (\ie \emph{Difficulty Calculator}, \emph{Curriculum Learner}, and \emph{Data Augmenter}), we conduct experiments to evaluate \emph{Humer} from these three perspectives.
For \emph{Difficulty Calculator}, we focus on the comparative experimental analysis of model-based and code-based difficulty calculators.
For \emph{Curriculum Learner}, we focus on the comparative experimental analysis of learning without curriculum and learning with curriculum.
For \emph{Data Augmenter}, we focus on the comparative experimental analysis of learning without data augmentation and learning with data augmentation.

\begin{table}[htbp]
  \centering
  \small
  \caption{Descriptions of used dataset in our experiments}
    \begin{tabular}{|c|cc|}
    \hline
          & \#Vulnerable Functions & \#Normal Functions \\
          \hline
    Qemu  & 7,479  & 1,0070 \\
    Ffmpeg & 4,981  & 4,788 \\
    Total & 12,460 & 14,858 \\
    \hline
    Ave \#LOC & 63    & 57 \\
    \hline
    \end{tabular}%
  \label{tab:dataset}%
\end{table}%

\subsection{Experimental Setup}
\subsubsection{Dataset}

\begin{table*}[htbp]
  \centering
  \small
  \caption{
  Experimental results of \emph{Humer} with model-based and code-based \emph{Difficulty Calculators}. DC denotes \emph{Difficulty Calculator}, CL denotes \emph{Curriculum Learner}. To complete model-based difficulty calculation, we first split the training set into $M$ subsets and train $M$ submodels. 
  To complete code-based difficulty calculation, we directly analyze the code complexity of all samples in the training set.
  After obtaining all samples' difficulty, we sort them and divide them into $N$ buckets for \emph{Curriculum Learner}.}
    \begin{tabular}{|c|c|c|c|c|c|c|c|c|c|c|c|c|c|}
    \hline
    \textbf{TokenCNN} & Baseline & \multicolumn{9}{c|}{Humer (model-based DC)}                           & \multicolumn{3}{c|}{Humer (code-based DC)} \\
    \hline
    $M$ for DC & None  & \multicolumn{3}{c|}{3} & \multicolumn{3}{c|}{5} & \multicolumn{3}{c|}{10} & None  & None  & None \\
    \hline
    $N$ for CL & None  & 3     & 5     & 10    & 3     & 5     & 10    & 3     & 5     & 10    & 3     & 5     & 10 \\
    \hline
    Accurary & 0.458 & 0.510 & \textbf{0.540} & 0.539 & 0.500 & 0.519 & 0.525 & 0.510 & 0.507 & 0.515 & 0.526 & 0.526 & 0.512 \\
    Recall & 0.510 & 0.673 & \textbf{0.677} & 0.662 & 0.665 & 0.642 & 0.620 & 0.621 & 0.670 & 0.628 & 0.636 & 0.606 & 0.571 \\
    Precision & 0.416 & 0.448 & \textbf{0.497} & 0.477 & 0.448 & 0.469 & 0.469 & 0.458 & 0.458 & 0.465 & 0.485 & 0.479 & 0.496 \\
    F1    & 0.458 & 0.538 & \textbf{0.573} & 0.554 & 0.535 & 0.542 & 0.534 & 0.527 & 0.544 & 0.535 & 0.550 & 0.535 & 0.531 \\
    \hline
    \textbf{VulDeePecker} & Baseline & \multicolumn{9}{c|}{Humer (model-based DC)}                           & \multicolumn{3}{c|}{Humer (code-based DC)} \\
    \hline
   Accurary & 0.532 & \textbf{0.635} & 0.554 & 0.580 & 0.582 & 0.547 & 0.587 & 0.554 & 0.564 & 0.556 & 0.579 & 0.544 & 0.551 \\
    Recall & 0.577 & 0.678 & 0.658 & 0.716 & 0.672 & \textbf{0.743} & 0.606 & 0.711 & 0.649 & 0.724 & 0.595 & 0.703 & 0.652 \\
    Precision & 0.472 & \textbf{0.585} & 0.501 & 0.524 & 0.525 & 0.497 & 0.542 & 0.495 & 0.511 & 0.502 & 0.530 & 0.471 & 0.498 \\
    F1    & 0.519 & \textbf{0.628} & 0.569 & 0.605 & 0.590 & 0.595 & 0.572 & 0.584 & 0.572 & 0.593 & 0.561 & 0.564 & 0.565 \\
    \hline
    \textbf{StatementGRU} & Baseline & \multicolumn{9}{c|}{Humer (model-based DC)}                           & \multicolumn{3}{c|}{Humer (code-based DC)} \\
    \hline
    Accurary & 0.537 & \textbf{0.649} & 0.572 & 0.600 & 0.595 & 0.565 & 0.600 & 0.567 & 0.581 & 0.569 & 0.585 & 0.580 & 0.576 \\
    Recall & 0.592 & 0.722 & 0.677 & 0.730 & 0.694 & \textbf{0.765} & 0.627 & 0.737 & 0.663 & 0.740 & 0.655 & 0.630 & 0.652 \\
    Precision & 0.488 & \textbf{0.572} & 0.518 & 0.539 & 0.538 & 0.507 & 0.552 & 0.512 & 0.528 & 0.513 & 0.524 & 0.518 & 0.534 \\
    F1    & 0.535 & \textbf{0.638} & 0.587 & 0.620 & 0.606 & 0.610 & 0.587 & 0.604 & 0.588 & 0.606 & 0.583 & 0.568 & 0.587 \\
    \hline
    \textbf{ASTGRU} & Baseline & \multicolumn{9}{c|}{Humer (model-based DC)}                           & \multicolumn{3}{c|}{Humer (code-based DC)} \\
    \hline
    Accurary & 0.580 & 0.627 & 0.617 & 0.650 & 0.621 & 0.648 & \textbf{0.671} & 0.651 & 0.631 & 0.638 & 0.637 & 0.639 & 0.636 \\
    Recall & 0.547 & 0.685 & 0.680 & 0.624 & \textbf{0.691} & 0.629 & 0.649 & 0.628 & 0.663 & 0.656 & 0.589 & 0.601 & 0.600 \\
    Precision & 0.540 & 0.567 & 0.564 & 0.622 & 0.572 & 0.616 & \textbf{0.642} & 0.614 & 0.584 & 0.596 & 0.545 & 0.567 & 0.549 \\
    F1    & 0.543 & 0.621 & 0.617 & 0.623 & 0.626 & 0.622 & \textbf{0.646} & 0.621 & 0.621 & 0.624 & 0.566 & 0.583 & 0.573 \\
    \hline
    \textbf{Devign} & Baseline & \multicolumn{9}{c|}{Humer (model-based DC)}                           & \multicolumn{3}{c|}{Humer (code-based DC)} \\
    \hline
    Accurary & 0.561 & 0.660 & \textbf{0.669} & 0.659 & 0.664 & 0.664 & 0.645 & 0.657 & 0.648 & 0.647 & 0.615 & 0.623 & 0.628 \\
    Recall & 0.572 & 0.678 & \textbf{0.703} & 0.682 & 0.681 & 0.669 & 0.688 & 0.673 & 0.657 & 0.662 & 0.621 & 0.615 & 0.629 \\
    Precision & 0.534 & 0.602 & 0.603 & 0.599 & 0.583 & 0.593 & 0.574 & 0.606 & \textbf{0.620} & 0.593 & 0.571 & 0.582 & 0.584 \\
    F1    & 0.553 & 0.638 & \textbf{0.649} & 0.638 & 0.628 & 0.629 & 0.626 & 0.638 & 0.638 & 0.626 & 0.595 & 0.598 & 0.606 \\
    \hline
    \end{tabular}%
  \label{tab:DC-results}%
\end{table*}%

To test \emph{Humer} on real-world vulnerability detection, we select a widely used open source vulnerability dataset \cite{zhou2019devign} which consists of 12,460 vulnerabilities and 14,858 normal functions.
A team of four professional security researchers spent 600 man-hours to construct it.
Functions in the dataset are mainly obtained from two open source software, namely qemu \cite{qemu} and ffmpeg \cite{ffmpeg}.
A total of 7,479 vulnerable functions and 10,070 and normal functions are analyzed from qemu, and a total of 4,981 vulnerable functions and 4,788 normal functions are obtained from ffmpeg.
The average number of \emph{lines of code} (LOC) of these vulnerabilities and normal functions are 63 and 57, respectively.

\subsubsection{Selected DL-based Vulnerability Detectors}
Since a program can be characterized by different code representations such as \emph{abstract syntactic tree} (AST) and \emph{control flow graph} (CFG).
We select five state-of-the-art DL-based vulnerability detection methods corresponding to five different code representations, respectively.
By this, we can achieve a more comprehensive evaluation of \emph{Humer}.
These five detectors\footnote{For convenience, we name the unnamed models based on their code representations (\eg token) and neural networks (\eg CNN).} are as follows:
\begin{itemize}
  \item \textbf{\emph{TokenCNN}} \cite{russell2018automated}: 
  It is a token-based vulnerability detector.
  After obtaining the tokens of a program by lexical analysis, \emph{TokenCNN} inputs them into a \emph{convolutional neural network} (CNN) to train a model for vulnerability detection.
  \item \textbf{\emph{VulDeePecker}} \cite{2018VulDeePecker}: 
  It is a slice-based vulnerability detector.
  After extracting the slices of a program by static analysis, \emph{VulDeePecker} inputs them into a \emph{bidirectional long short-term memory} (BLSTM) network to train a model for vulnerability detection
  \item \textbf{\emph{StatementGRU}} \cite{lin2019statementgru}: 
  It is a statement-based vulnerability detector.
  \emph{StatementGRU} treats the statements of a program as sentences and directly inputs them into a \emph{bidirectional gated recurrent unit} (BGRU) network to train a model for vulnerability detection.
  \item \textbf{\emph{ASTGRU}} \cite{feng2020astgru}: 
  It is a tree-based vulnerability detector.
  After obtaining the AST of a program by static analysis, \emph{ASTGRU} first applies preorder traversal search algorithm to transform the AST into a sequence and then inputs the sequence into a \emph{bidirectional gated recurrent unit} (BGRU) network to train a model for vulnerability detection.
  \item \textbf{\emph{Devign}} \cite{zhou2019devign}: 
  It is a graph-based vulnerability detector.
  After extracting the \emph{abstract syntactic tree} (AST), \emph{control flow graph} (CFG), \emph{data flow graph} (DFG), and \emph{natural code sequence} (NSC) of a program.
  \emph{Devign} inputs them into a \emph{graph neural networks} (GNN) to train a model for vulnerability detection.
\end{itemize}

\subsubsection{Implementations}

We run all experiments on a server with 32 cores of CPU and a GTX 2080Ti GPU. 
For the dataset, we first randomly divide them into ten subsets, then the eight subsets are used for training, the other one subset is used for validating, and the last subset is used for testing. 
\subsubsection{Metrics}

We adopt the widely used metrics to evaluate the detection results of \emph{Humer} as follows:
\begin{itemize}
\item \emph{True Positive} (TP): the number of samples correctly predicted as vulnerable
\item \emph{True Negative} (TN): the number of samples correctly predicted as normal
\item \emph{False Positive} (FP): the number of samples incorrectly classified as vulnerable
\item \emph{False Negative} (FN): the number of samples incorrectly classified as normal
\item Accuracy=(TP+TN)/(TP+TN+FP+FN)
\item Recall=TP/(TP+FN)
\item Precision=TP/(TP+FP)
\item F1=2$*$Precision$*$Recall/(Precision+Recall)
\end{itemize}


\begin{figure*}
\centering
\subfigure{
\begin{minipage}[t]{0.23\textwidth}
\centering
\includegraphics[width=\textwidth]{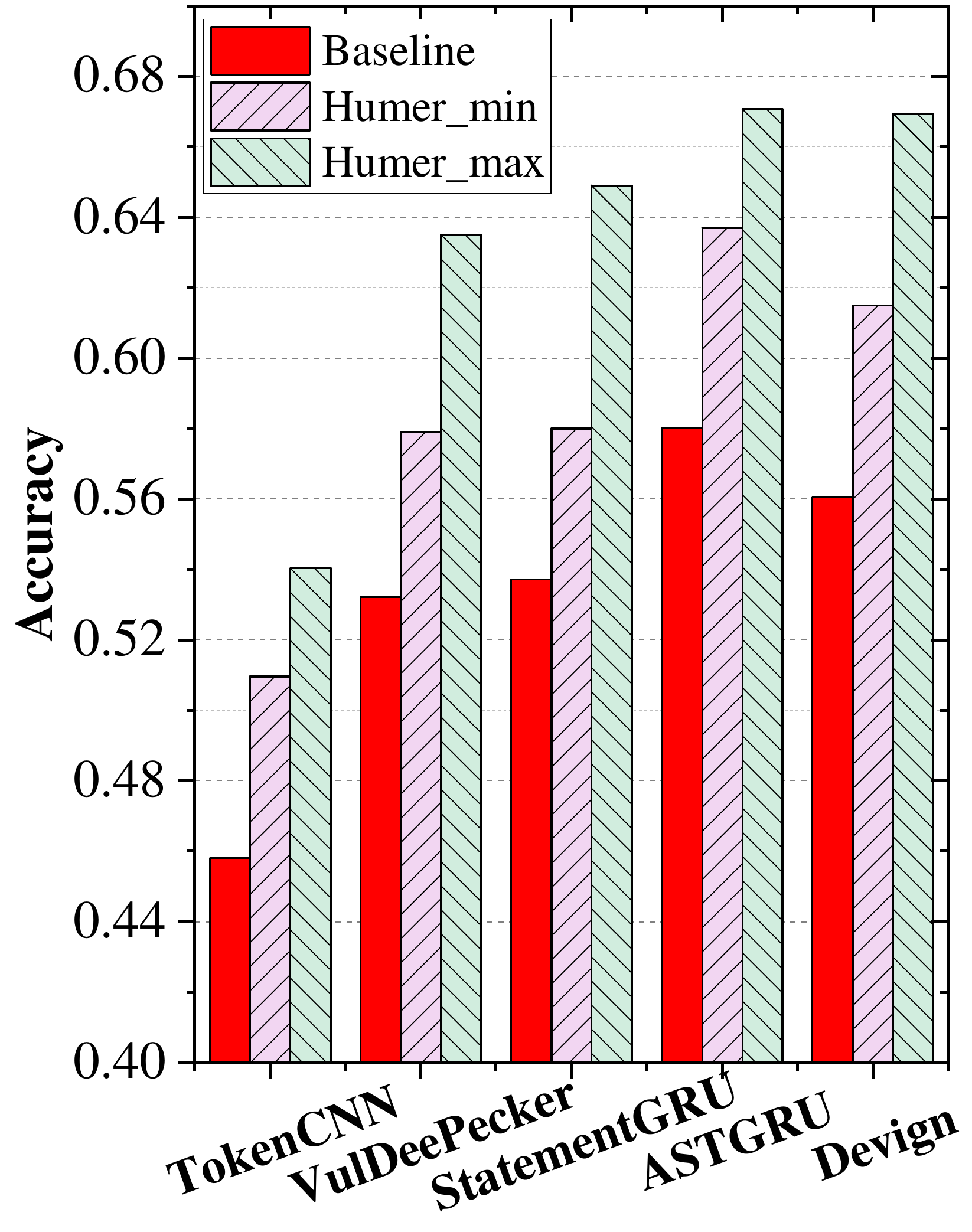}
\end{minipage}
}
\subfigure{
\begin{minipage}[t]{0.23\textwidth}
\centering
\includegraphics[width=\textwidth]{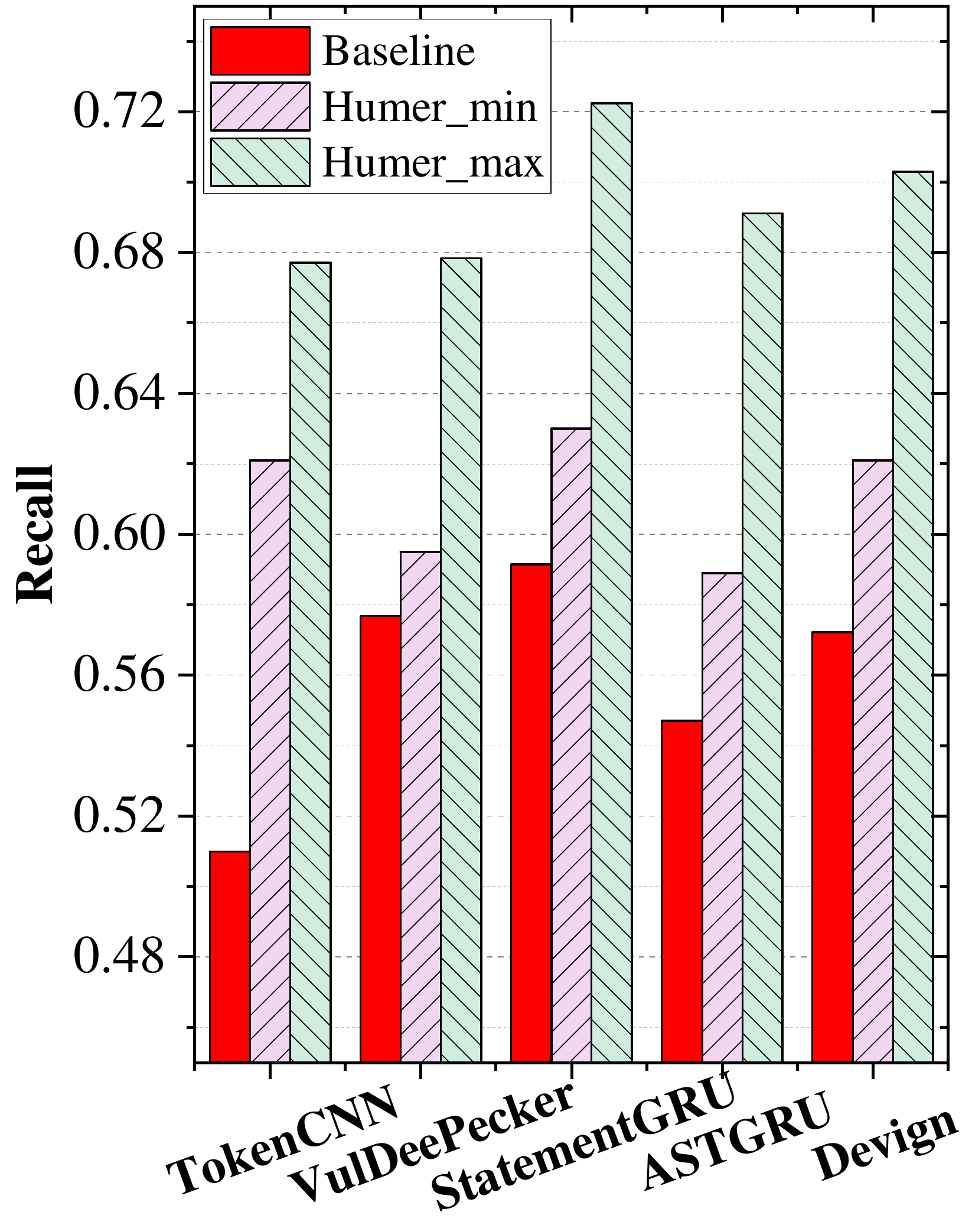}
\end{minipage}
}
\subfigure{
\begin{minipage}[t]{0.23\textwidth}
\centering
\includegraphics[width=\textwidth]{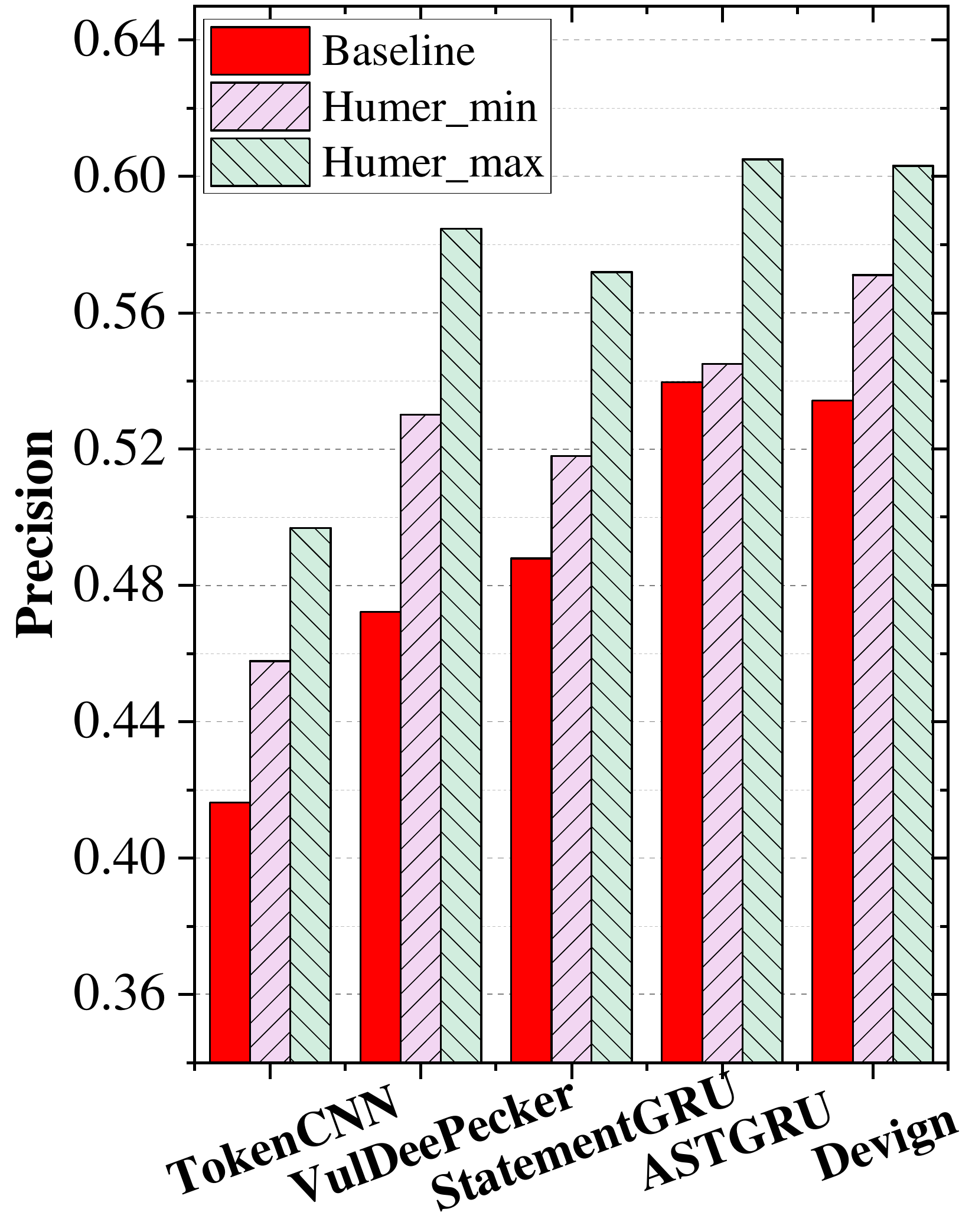}
\end{minipage}
}
\subfigure{
\begin{minipage}[t]{0.23\textwidth}
\centering
\includegraphics[width=\textwidth]{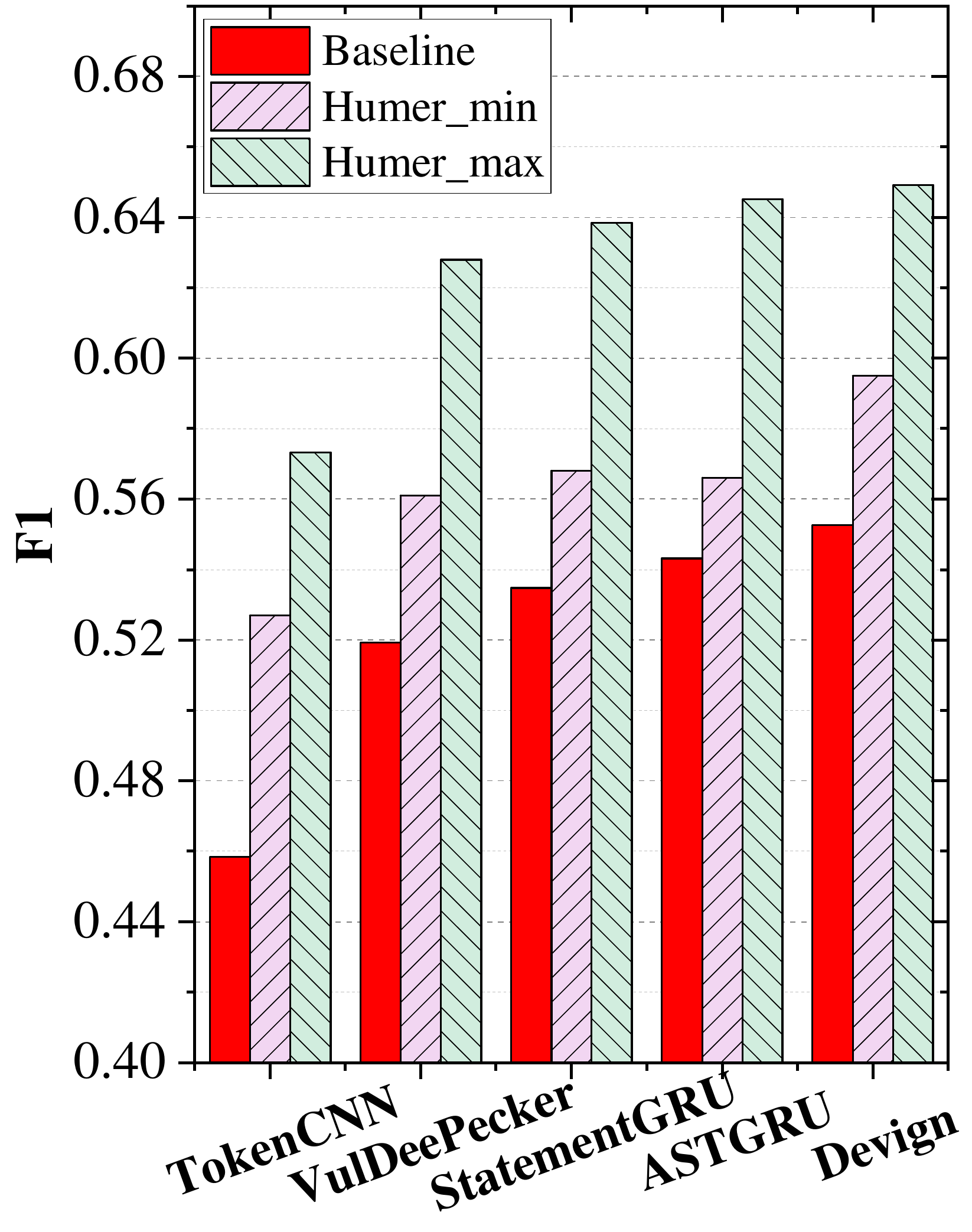}
\end{minipage}
}
\caption{The comparative results of DL-based vulnerability detectors (\ie \emph{TokenCNN}, \emph{VulDeePecker}, \emph{StatementGRU}, \emph{ASTGRU}, and \emph{Devign}) with and without \emph{Humer}}
\label{fig:CL-results}
\end{figure*}

\subsection{For Difficulty Calculator}

As aforementioned, we not only design a \emph{Difficulty Calculator} from the perspective of the model itself, but also implement another alternative \emph{Difficulty Calculator} by directly analyzing the code complexity of vulnerability samples.
Therefore, in this part, we focus on evaluating the detection effectiveness of \emph{Humer} with these two \emph{Difficulty Calculators}.
For model-based \emph{Difficulty Calculator}, we first split the training set into $M$ subsets and train $M$ submodels.
These submodels are used to evaluate the difficulty of samples in the training set.
In our experiments, we select three different $M$ (\ie 3, 5, and 10) to commence model-based difficulty calculation.
For code-based \emph{Difficulty Calculator}, we do not split the training set but directly analyze the code complexity of all samples.
After computing the difficulty, we sort these samples according to their difficulty scores and divide them into $N$ buckets uniformly.
In this way, we can train the model in a meaningful order, that is, learn from the simplest bucket and then transition to difficult buckets.
We also chose three different $N$ (\ie 3, 5, and 10) to train the model.

Table \ref{tab:DC-results} presents the detailed results of \emph{Humer} with model-based and code-based \emph{Difficulty Calculators}.
Through these results, we obtain several findings.
The first is that when \emph{Humer} adopts model-based \emph{Difficulty Calculator}, the detection effectiveness is more improved.
For example, when training \emph{VulDeePecker}, if \emph{Humer} uses model-based calculator to compute the difficulty, the F1 can be increased to 62.8\%, but it can only be increased to 56.5\% when adopts code-based calculator.
It is reasonable because we humans think that complex and difficult samples are not necessarily difficult for the model. 
In fact, our training object is a model, not humans. 
If the difficulty is calculated from the perspective of the model itself, the difficulty will be more suitable for the model, making the training of the model more meaningful.

The second is that when different $M$ and $N$ are selected, the improved effectiveness is also different.
This is normal, because when different $M$ is used (\ie the training set is divided into different numbers of subsets), the difficulty of the calculated sample may be different due to the different number of submodels.
This can affect the subsequent difficulty sorting, and then affect the final training effectiveness.
When using different $N$ (\ie the sorted samples are divided into different numbers of buckets), the number of samples in each bucket will be different, and different numbers of training samples will make the final training effectiveness different.
In fact, we also observe that the overall effectiveness is generally best when it is divided into three subsets (\ie $M$ = 3) to calculate the difficulty.
When the number of subsets increases, the number of vulnerabilities in each subset will decrease. 
If the number of vulnerabilities is small, after training the submodel, the submodel itself will not learn much knowledge of vulnerabilities, which may lead to errors when using these submodels to evaluate the difficulty.
We will choose more parameters in the future to find the most suitable combination.

\subsection{For Curriculum Learner}

In this part, we focus on the comparative analysis of DL-based vulnerability detectors with and without \emph{Humer}.
A shown in Table \ref{tab:DC-results}, since we select three different $M$ and three different $N$ to test \emph{Humer}, we can obtain a total of twelve experimental results (\ie nine experimental results for model-based \emph{Difficulty Calculator} and three experimental results for code-based \emph{Difficulty Calculator}) for each model.
To present it more clearly, we only show the lowest and highest results among these twelve results in Figure \ref{fig:CL-results}.

\begin{table}[htbp]
  \centering
  \small
  \caption{The improvement of DL-based vulnerability detectors (\ie \emph{TokenCNN}, \emph{VulDeePecker}, \emph{StatementGRU}, \emph{ASTGRU}, and \emph{Devign}) after using \emph{Humer} to train them}
    \begin{tabular}{|c|c|c|c|c|}
    \hline
     & Accurary & Recall & Precision & F1 \\
     \hline
    TokenCNN & 8.2\% & 16.7\% & 8.1\% & 11.5\% \\
    VulDeePecker & 10.3\% & 10.1\% & 11.3\% & 10.9\% \\
    StatementGRU & 11.2\% & 13.0\% & 8.4\% & 10.3\% \\
    ASTGRU & 9.1\% & 10.2\% & 10.2\% & 10.3\% \\
    Devign & 10.8\% & 13.1\% & 6.9\% & 9.6\% \\
    \hline
    \textbf{Average} & \textbf{9.9\%} & \textbf{12.6\%} & \textbf{9.0\%} & \textbf{10.5\%} \\
    \hline
    \end{tabular}%
  \label{tab:CL-improvement}%
\end{table}%

\begin{figure*}
\centering
\subfigure{
\begin{minipage}[t]{0.23\textwidth}
\centering
\includegraphics[width=\textwidth]{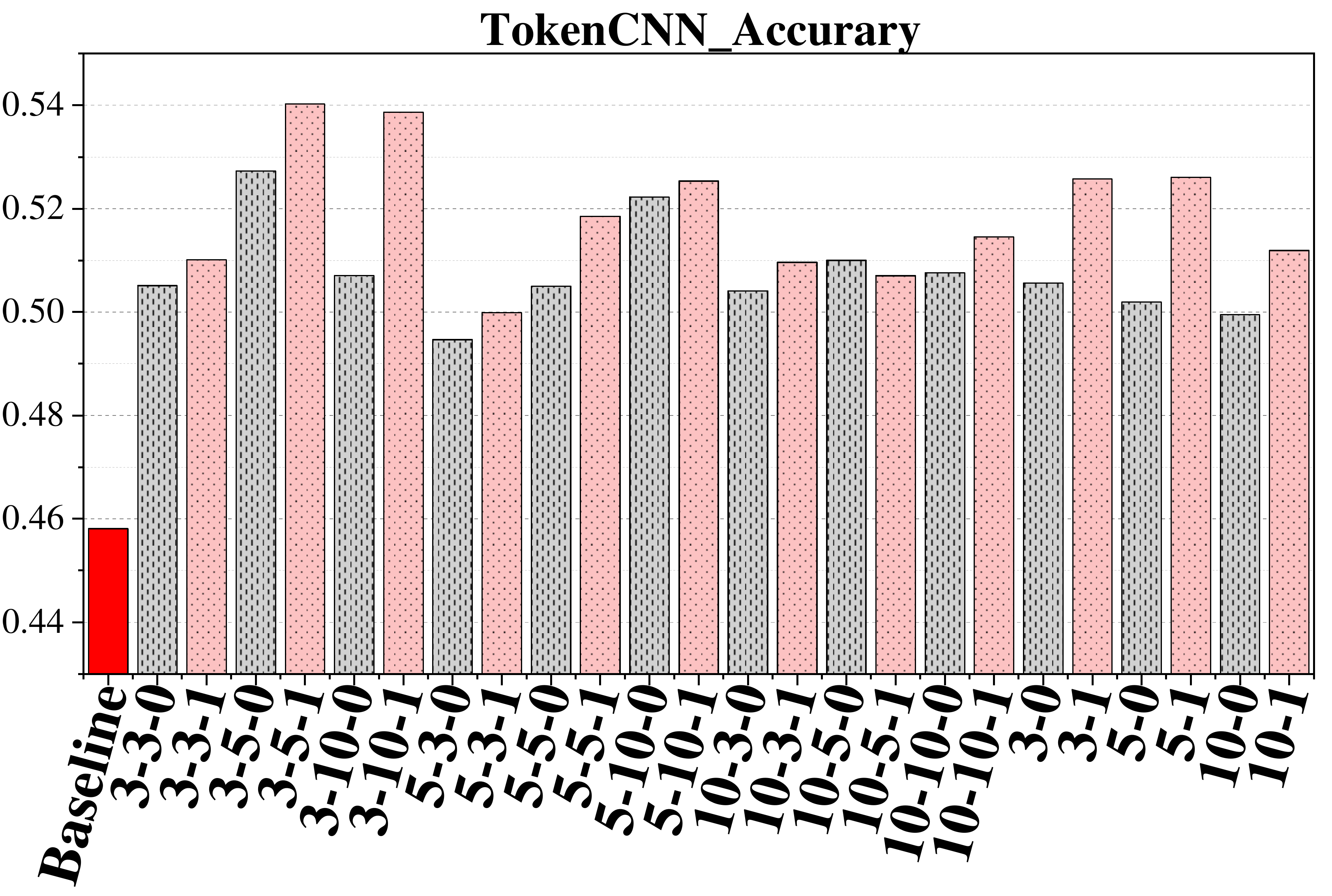}
\end{minipage}
}
\subfigure{
\begin{minipage}[t]{0.23\textwidth}
\centering
\includegraphics[width=\textwidth]{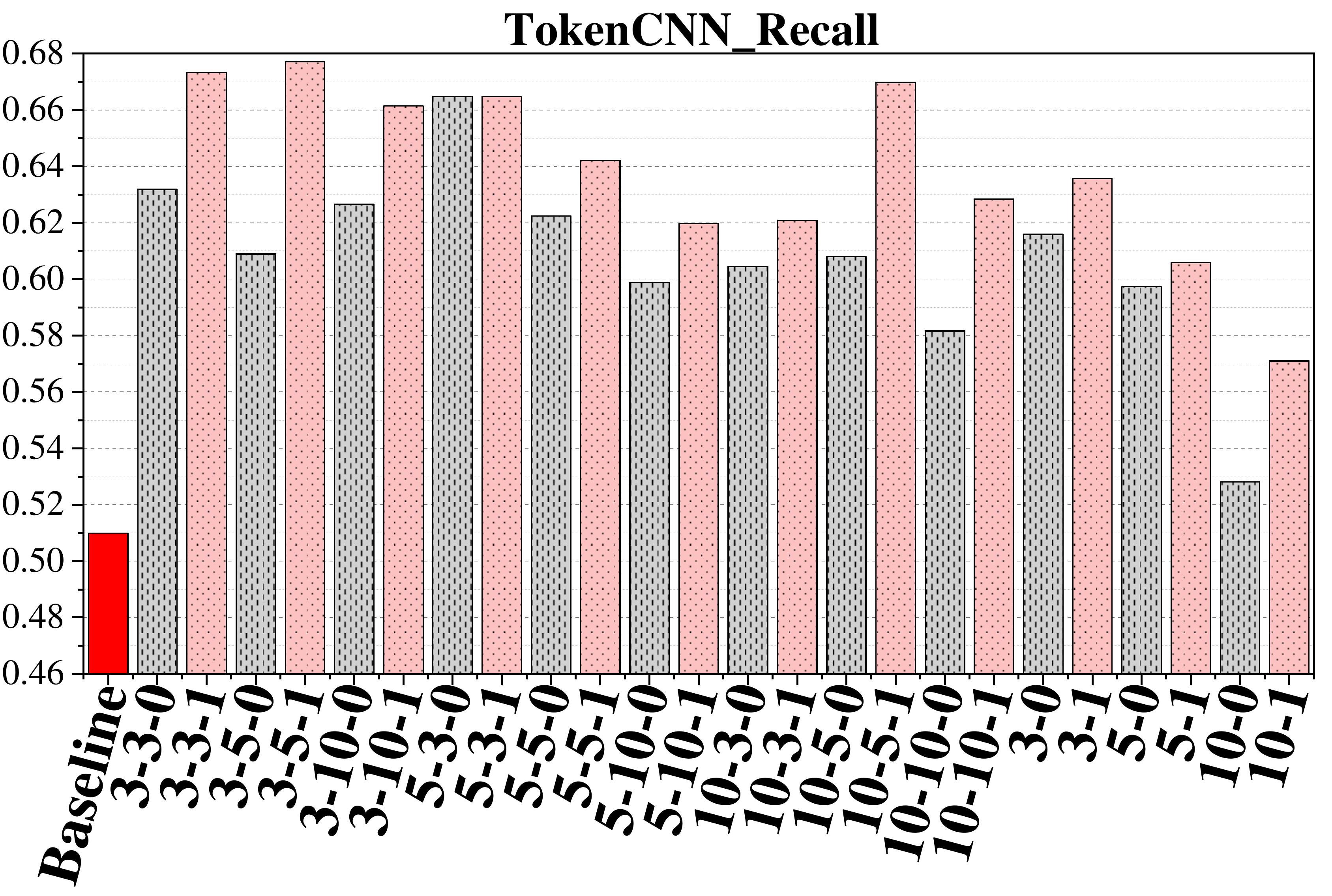}
\end{minipage}
}
\subfigure{
\begin{minipage}[t]{0.23\textwidth}
\centering
\includegraphics[width=\textwidth]{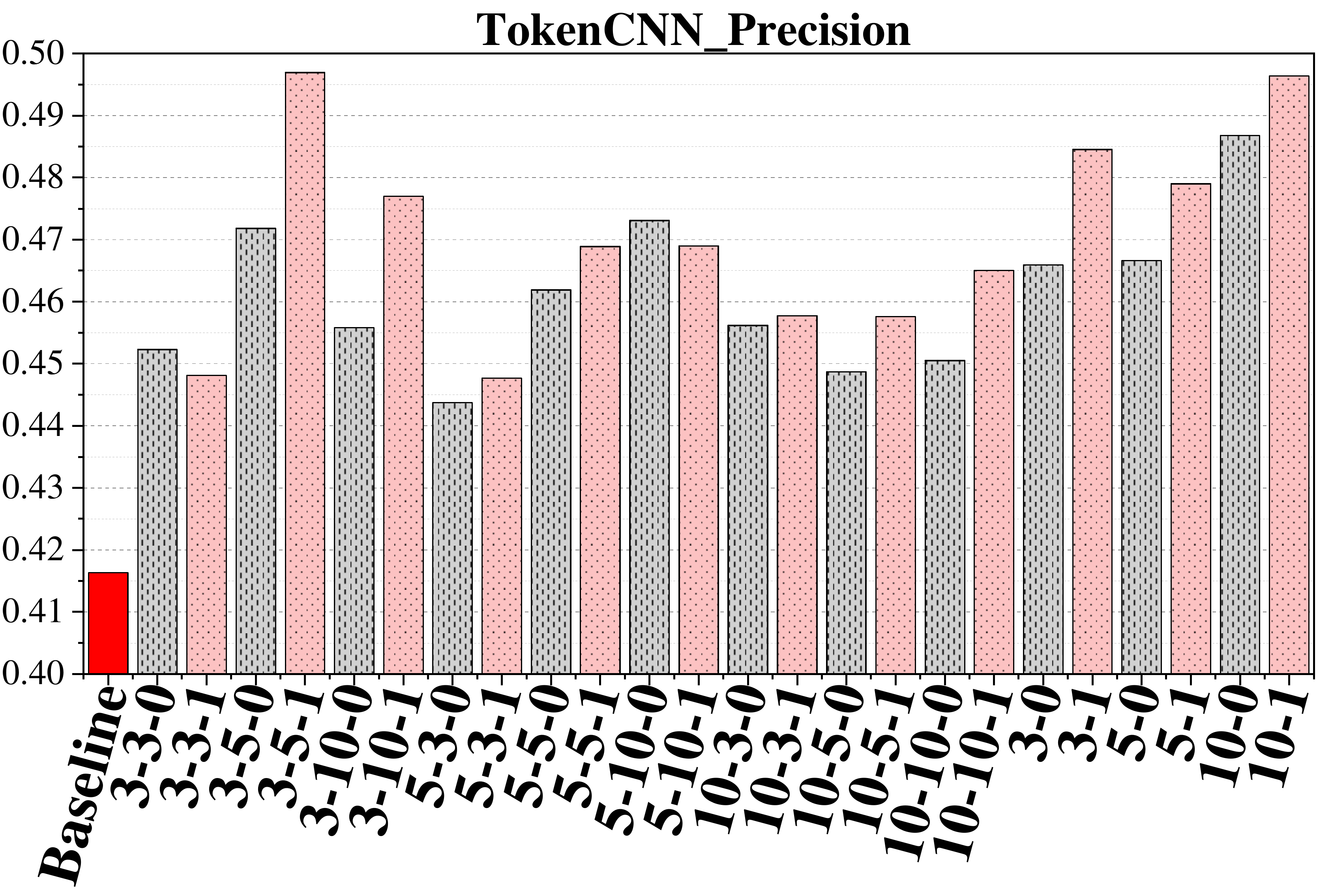}
\end{minipage}
}
\subfigure{
\begin{minipage}[t]{0.23\textwidth}
\centering
\includegraphics[width=\textwidth]{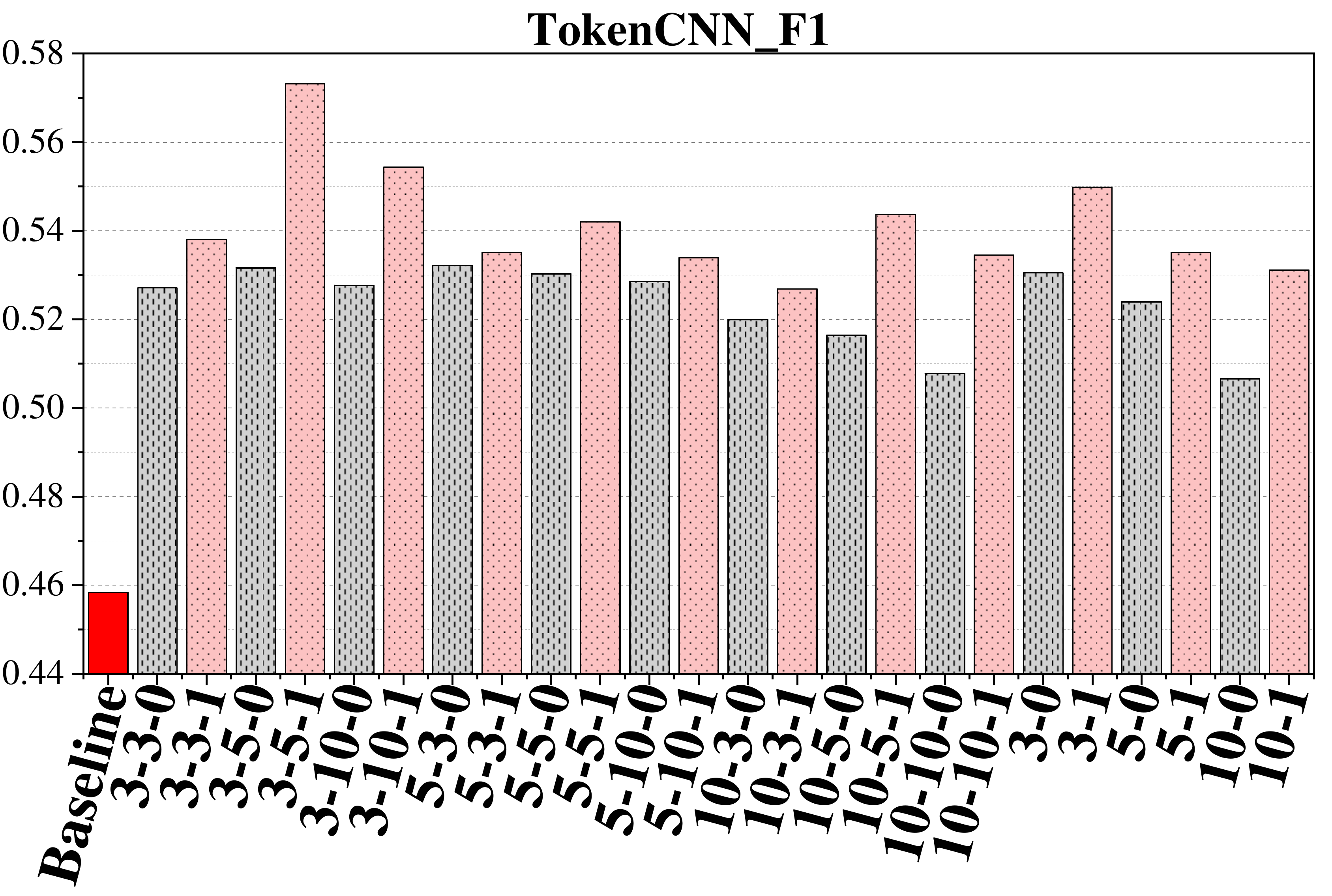}
\end{minipage}
}
\subfigure{
\begin{minipage}[t]{0.23\textwidth}
\centering
\includegraphics[width=\textwidth]{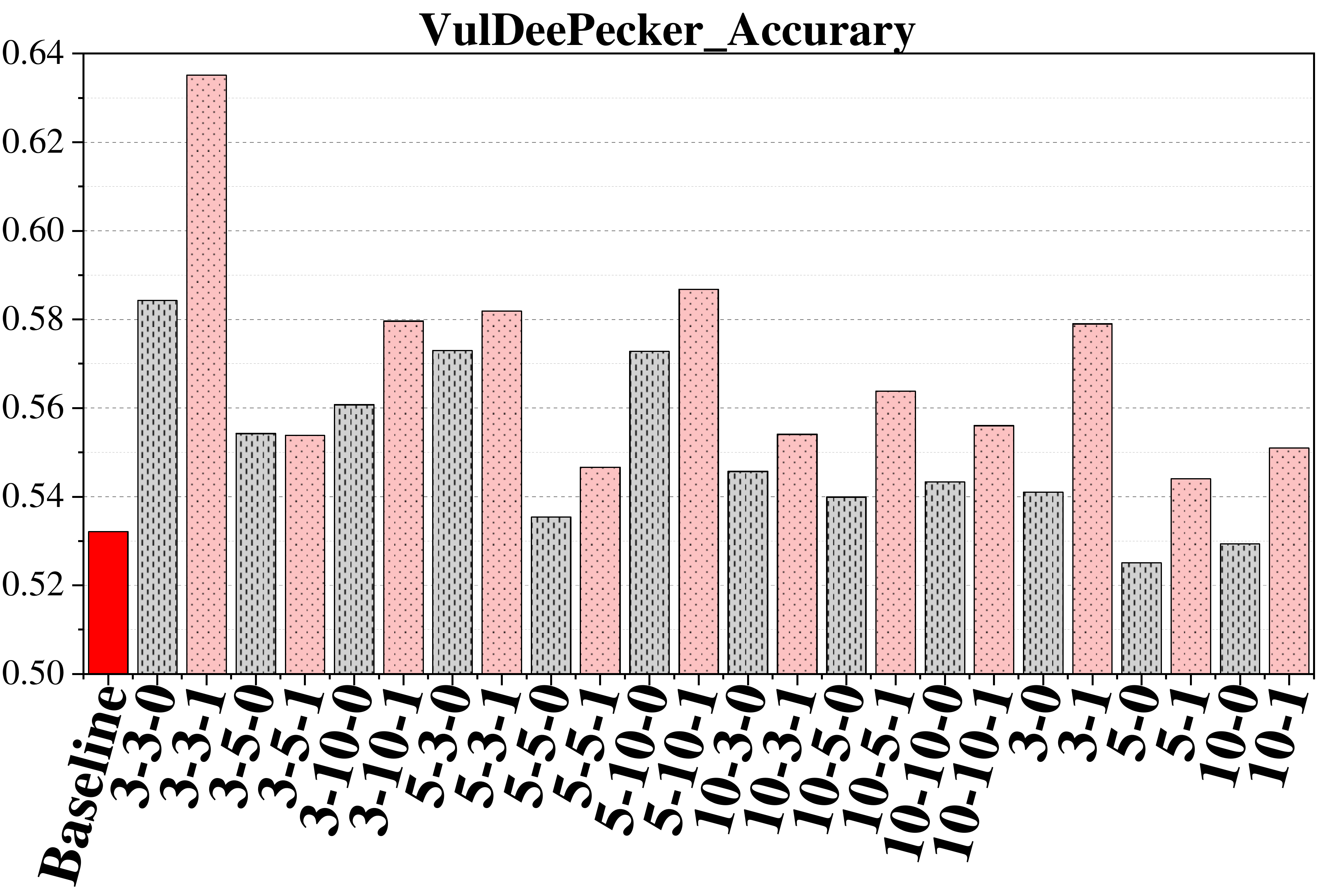}
\end{minipage}
}
\subfigure{
\begin{minipage}[t]{0.23\textwidth}
\centering
\includegraphics[width=\textwidth]{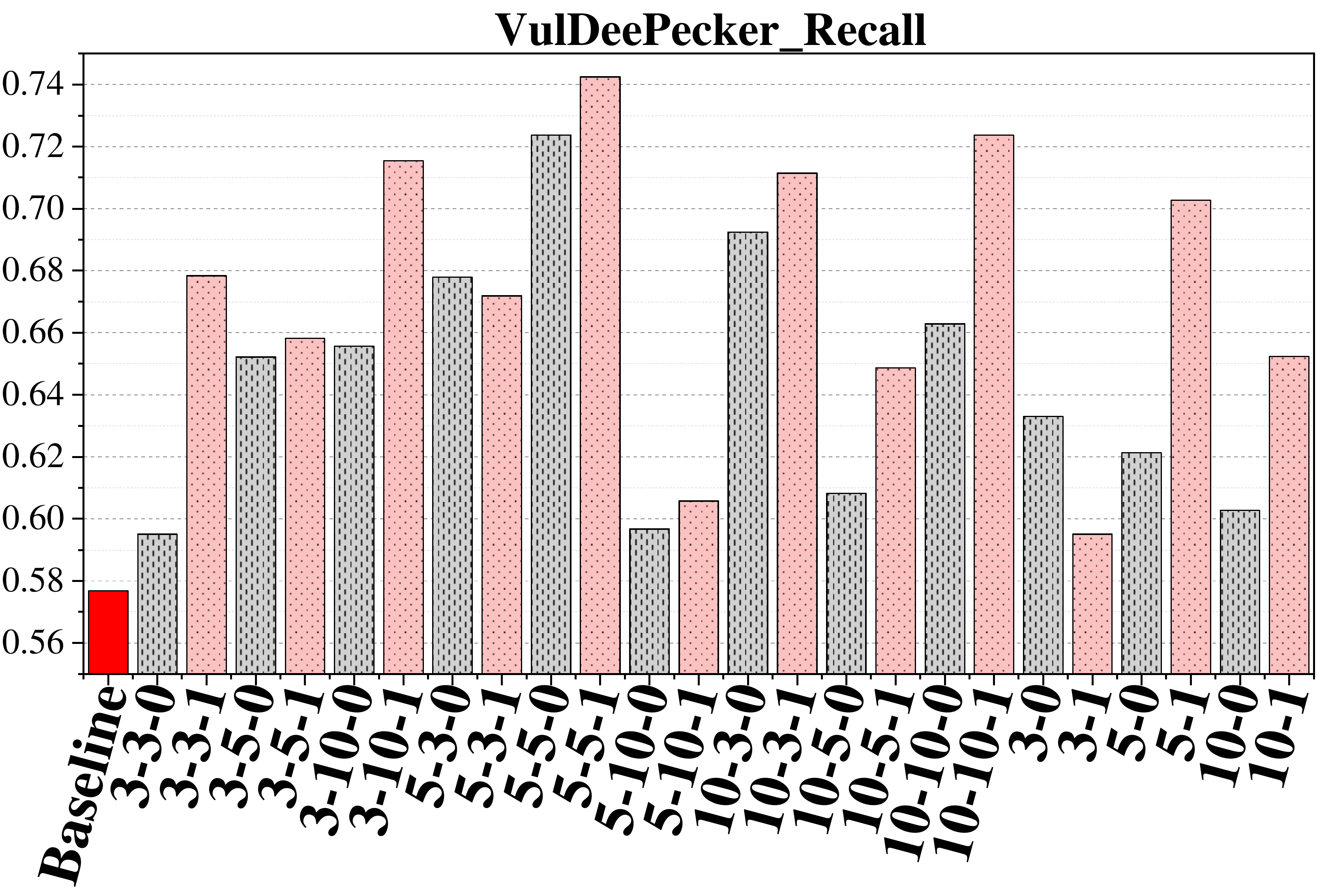}
\end{minipage}
}
\subfigure{
\begin{minipage}[t]{0.23\textwidth}
\centering
\includegraphics[width=\textwidth]{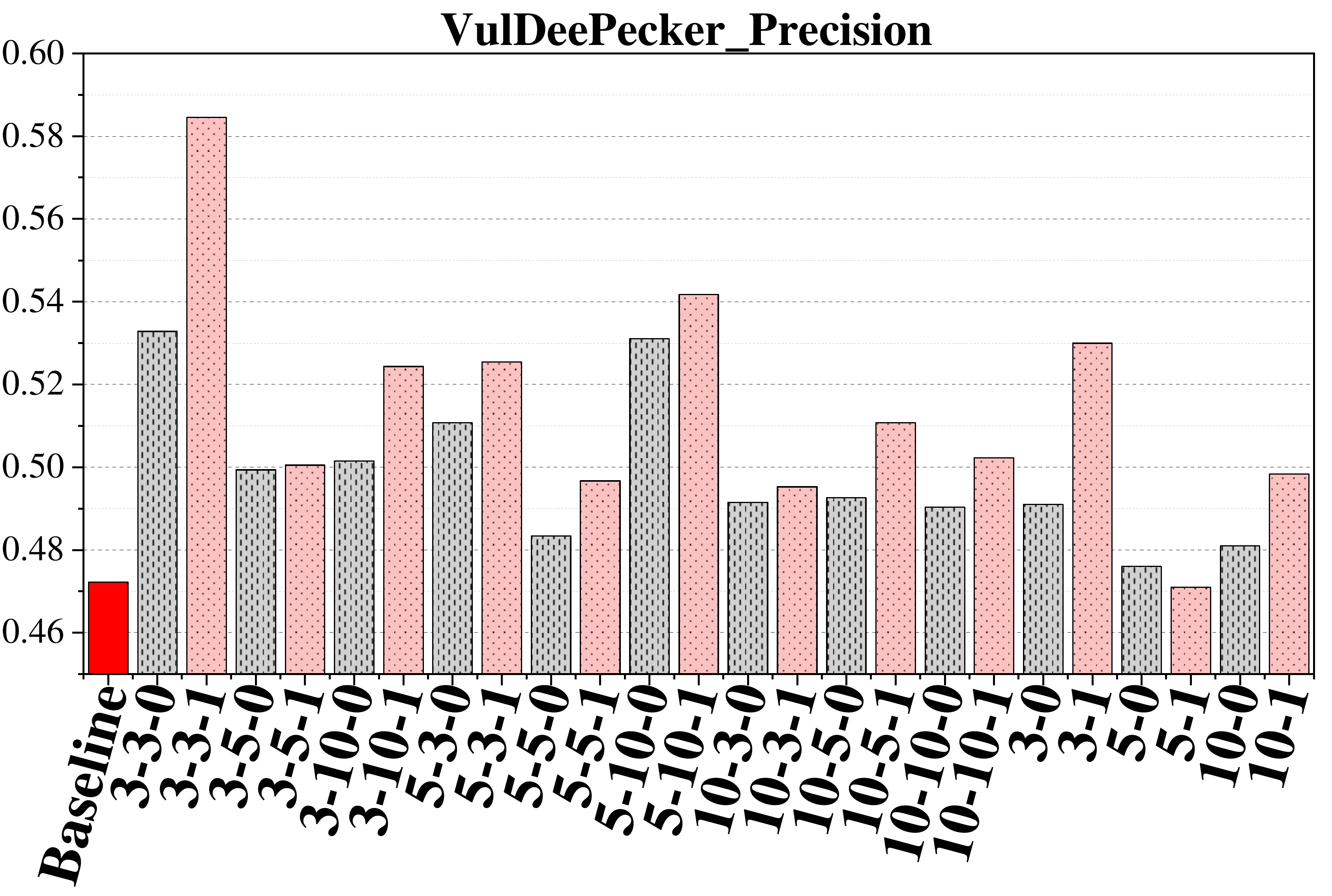}
\end{minipage}
}
\subfigure{
\begin{minipage}[t]{0.23\textwidth}
\centering
\includegraphics[width=\textwidth]{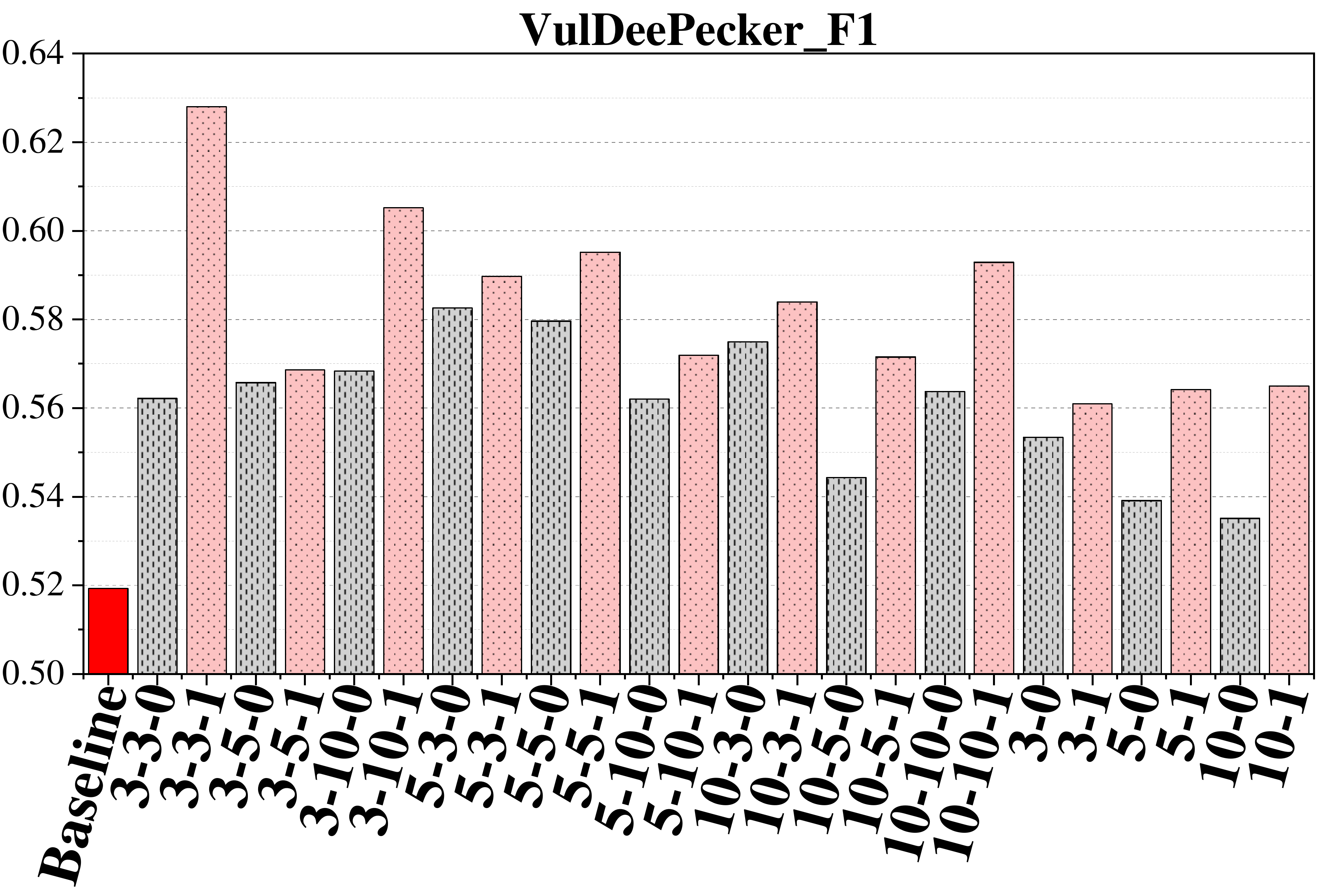}
\end{minipage}
}
\subfigure{
\begin{minipage}[t]{0.23\textwidth}
\centering
\includegraphics[width=\textwidth]{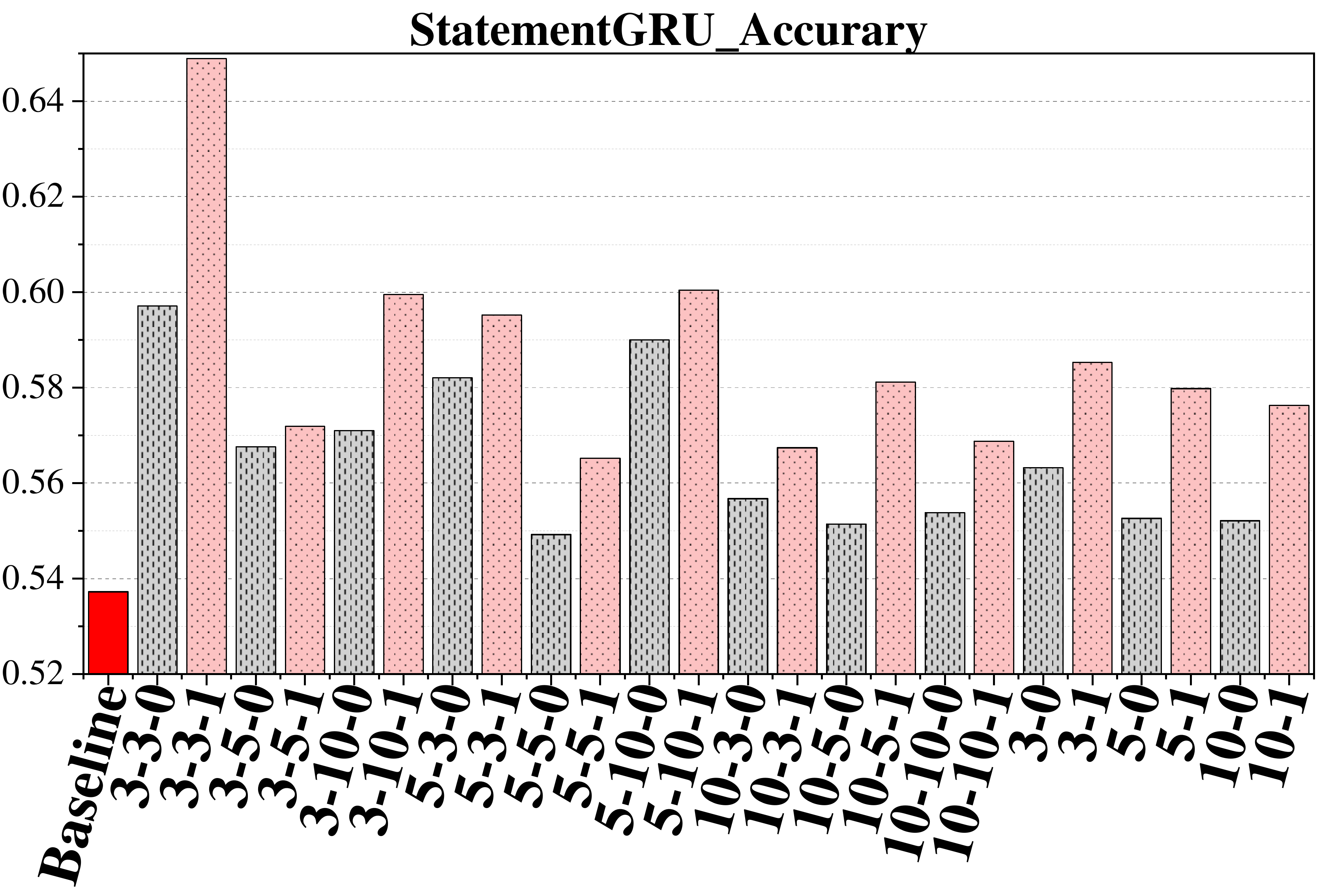}
\end{minipage}
}
\subfigure{
\begin{minipage}[t]{0.23\textwidth}
\centering
\includegraphics[width=\textwidth]{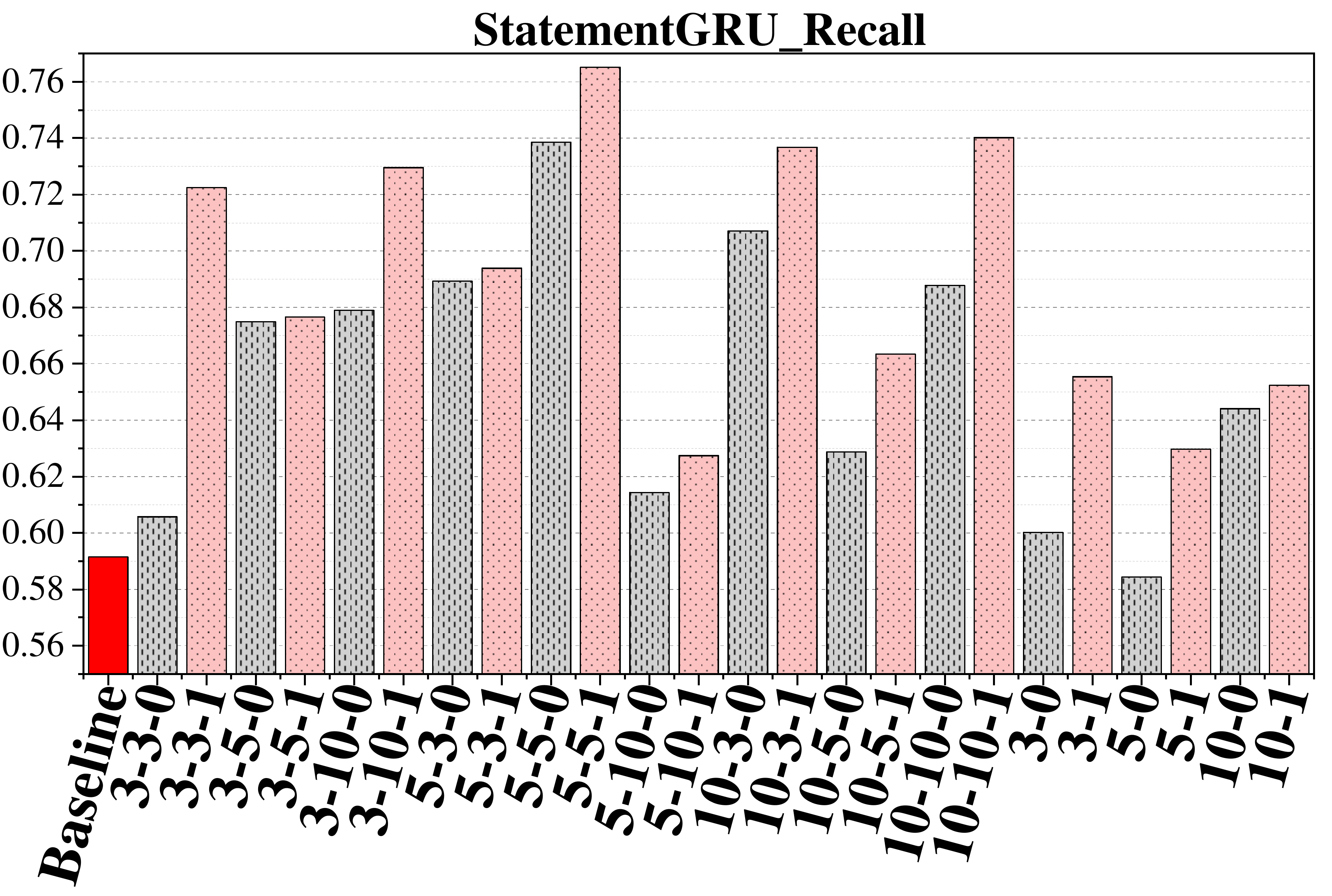}
\end{minipage}
}
\subfigure{
\begin{minipage}[t]{0.23\textwidth}
\centering
\includegraphics[width=\textwidth]{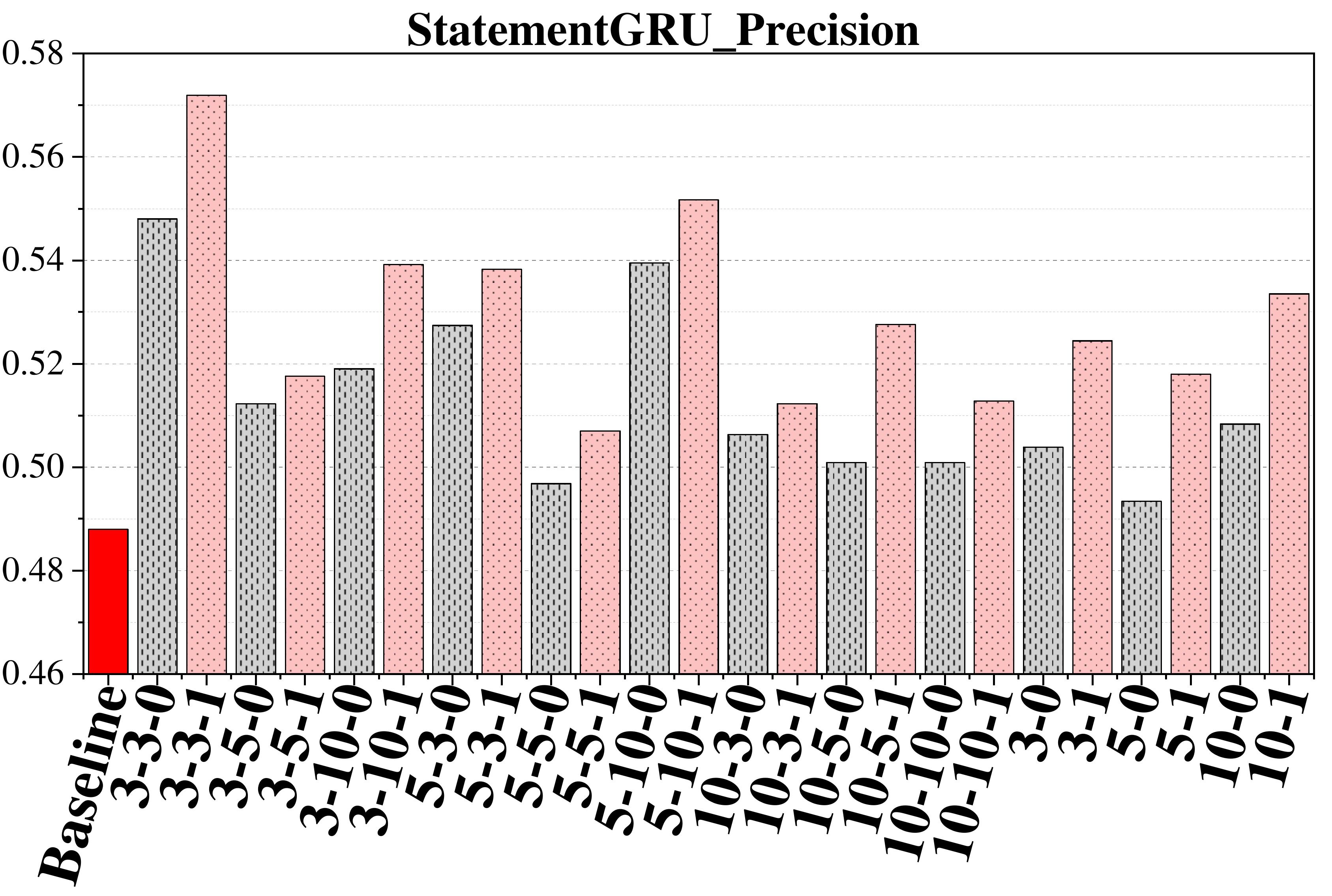}
\end{minipage}
}
\subfigure{
\begin{minipage}[t]{0.23\textwidth}
\centering
\includegraphics[width=\textwidth]{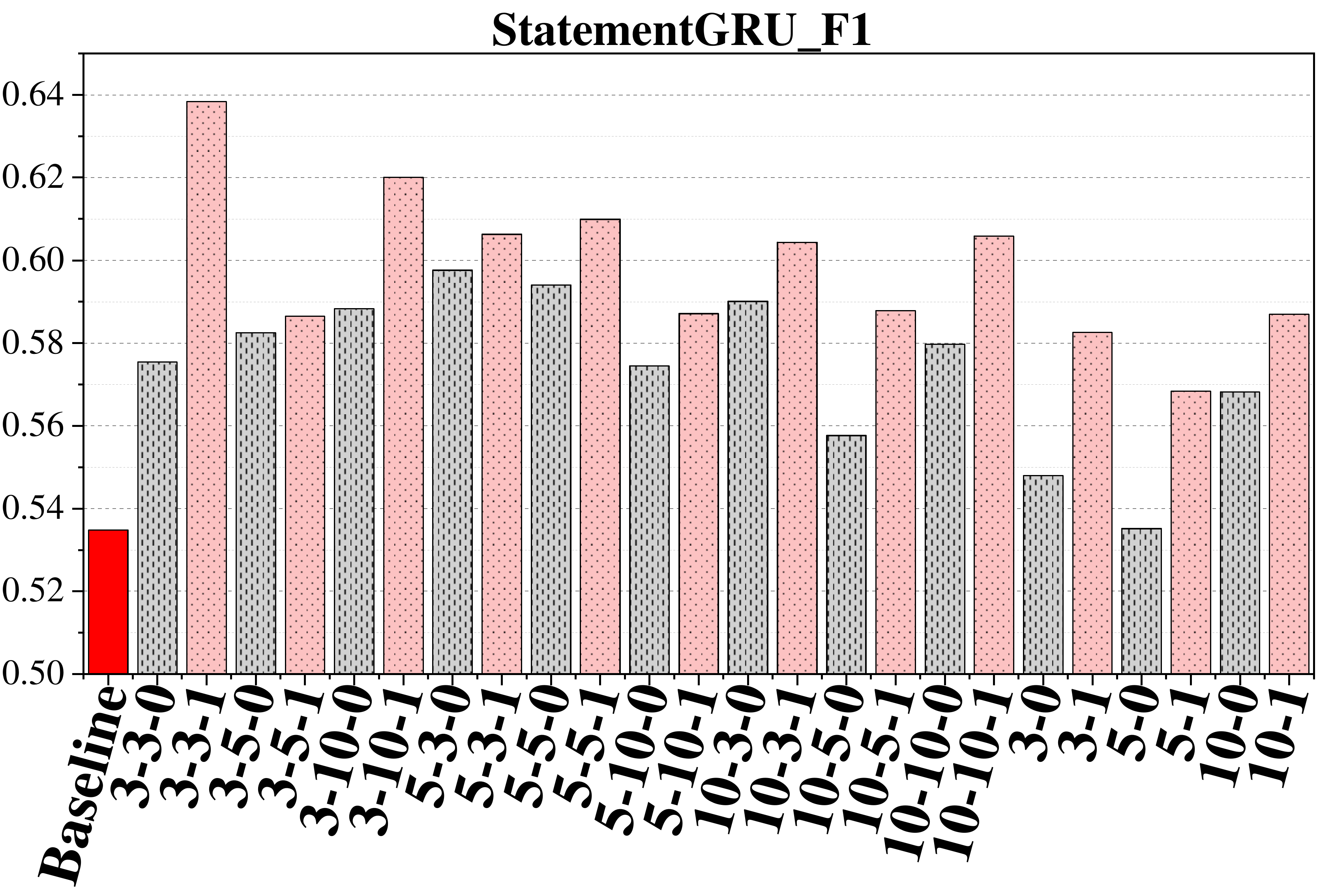}
\end{minipage}
}
\subfigure{
\begin{minipage}[t]{0.23\textwidth}
\centering
\includegraphics[width=\textwidth]{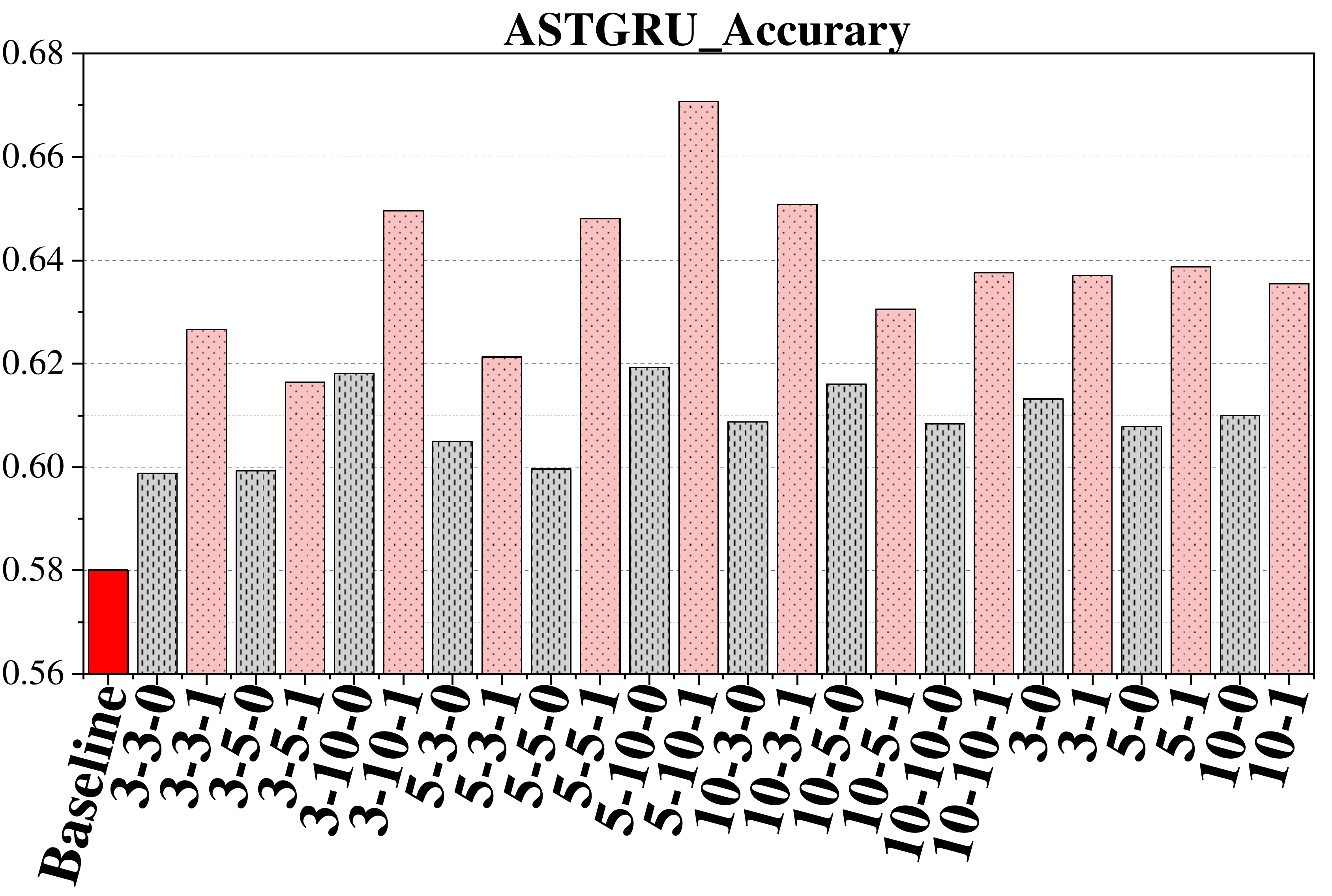}
\end{minipage}
}
\subfigure{
\begin{minipage}[t]{0.23\textwidth}
\centering
\includegraphics[width=\textwidth]{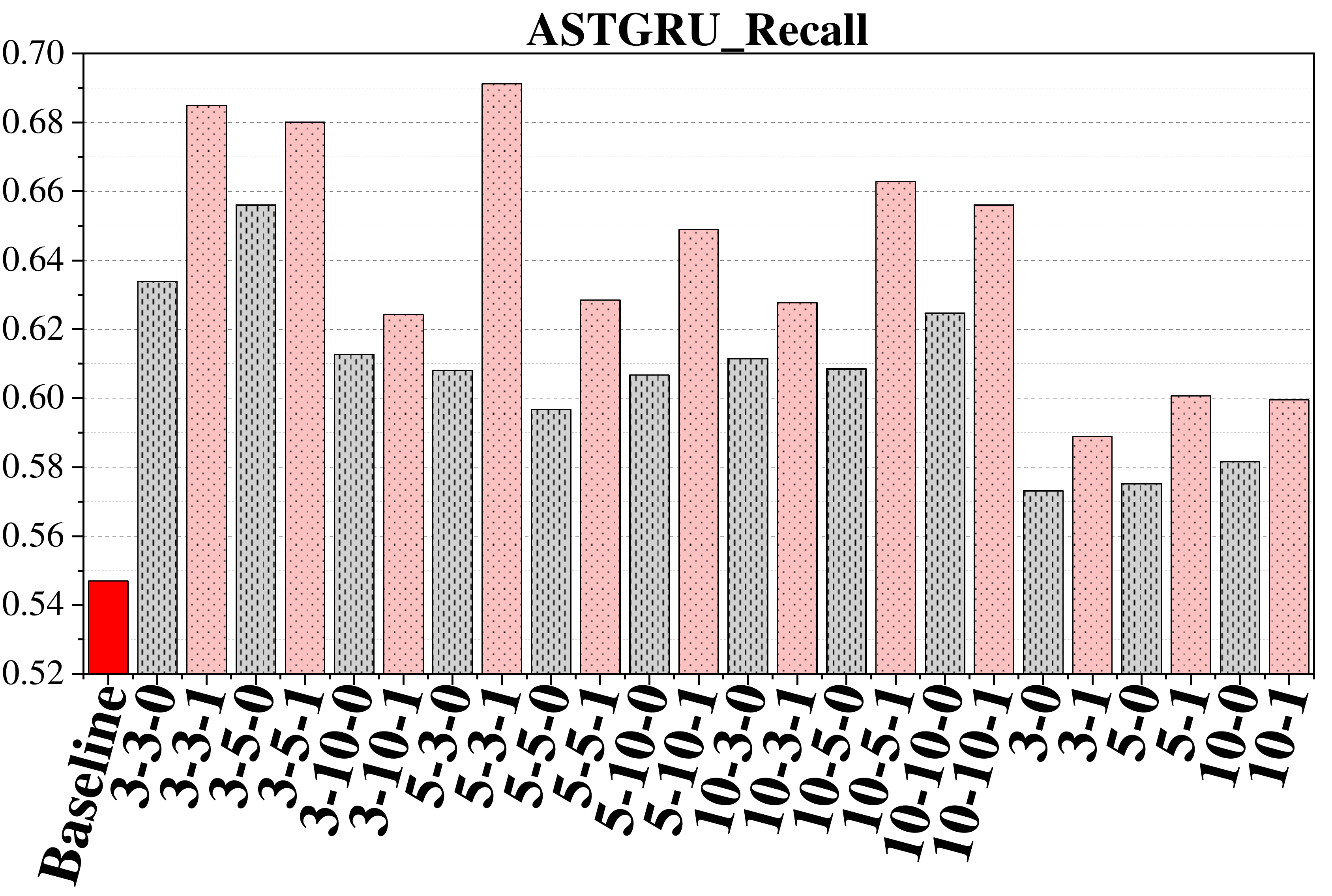}
\end{minipage}
}
\subfigure{
\begin{minipage}[t]{0.23\textwidth}
\centering
\includegraphics[width=\textwidth]{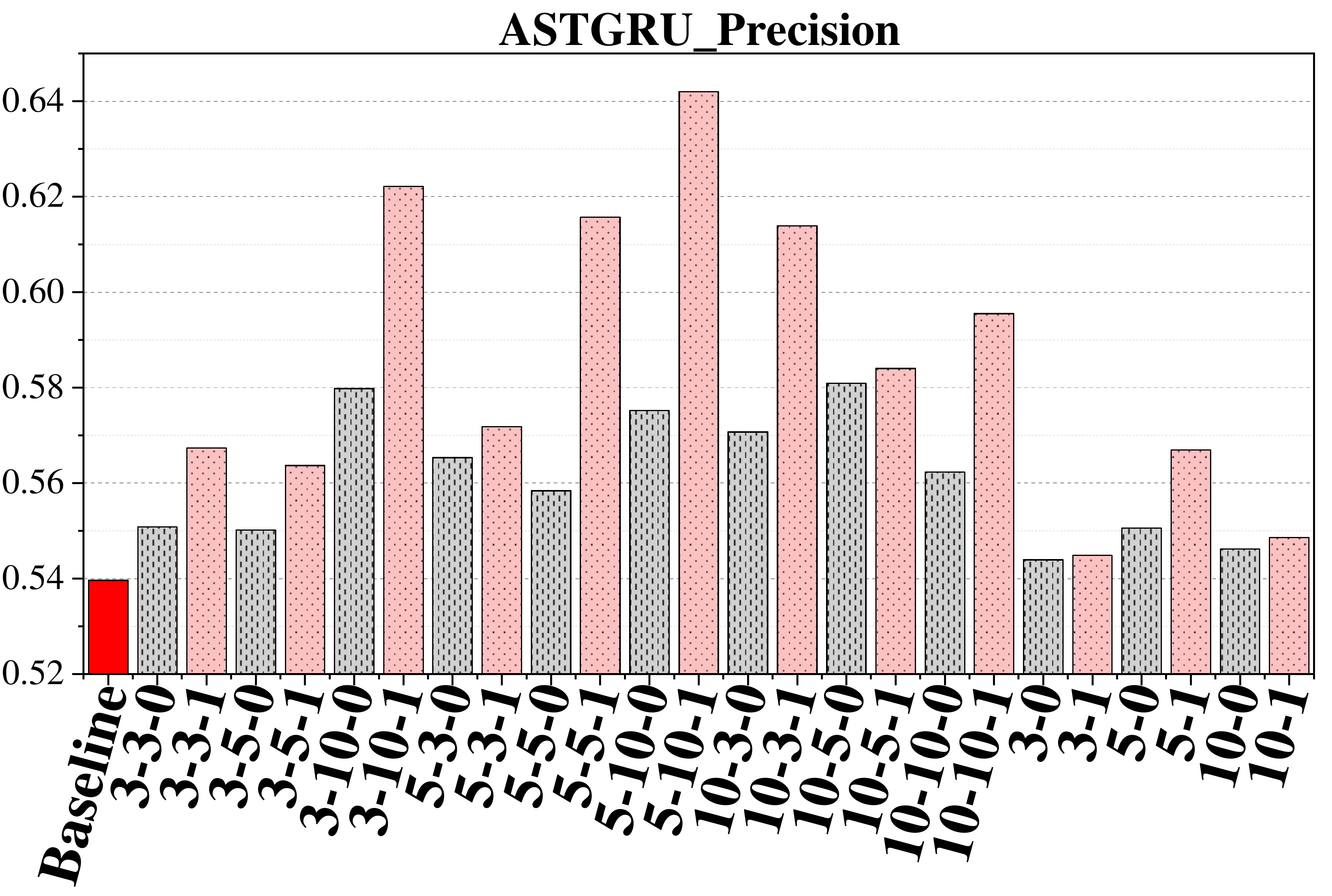}
\end{minipage}
}
\subfigure{
\begin{minipage}[t]{0.23\textwidth}
\centering
\includegraphics[width=\textwidth]{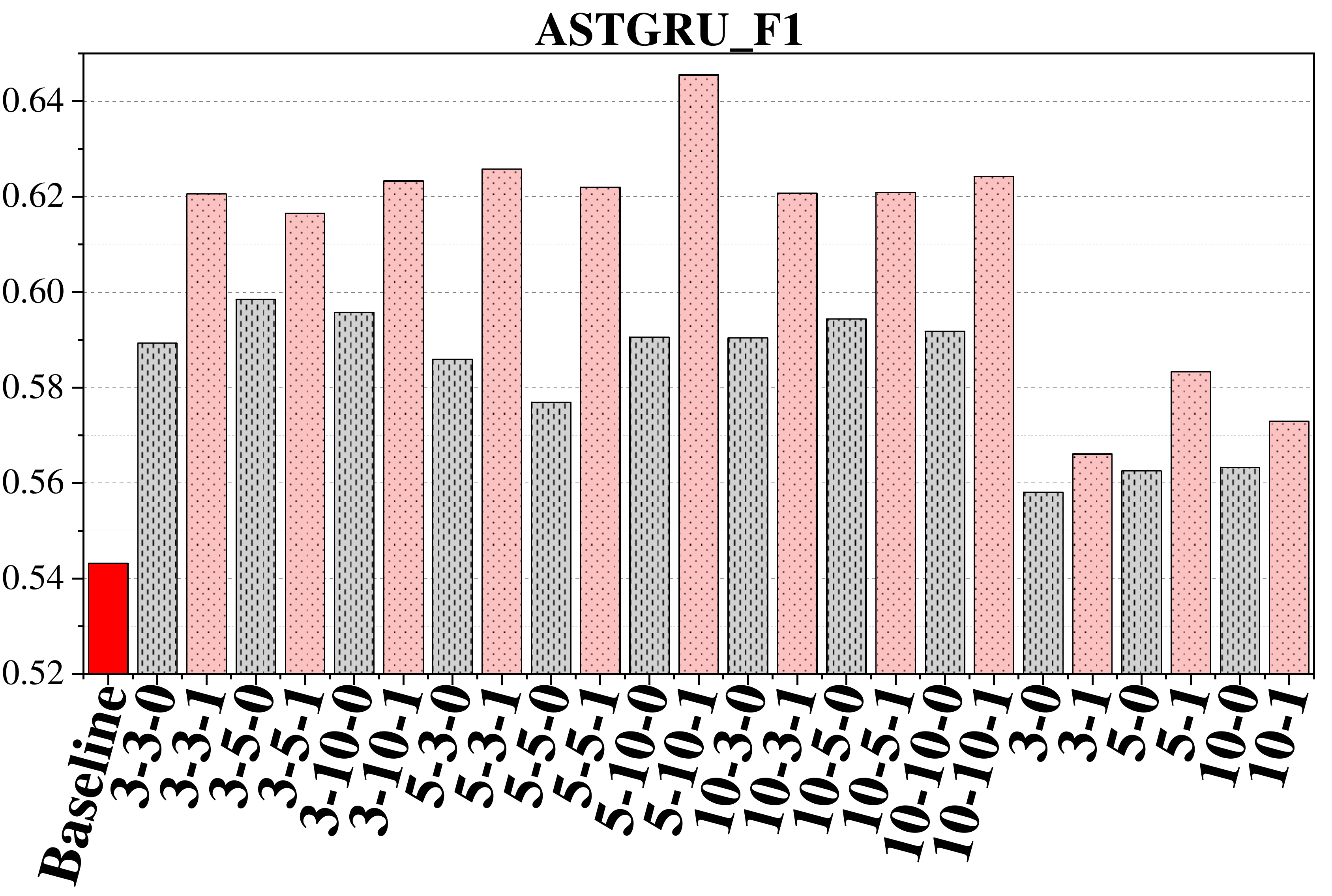}
\end{minipage}
}
\subfigure{
\begin{minipage}[t]{0.23\textwidth}
\centering
\includegraphics[width=\textwidth]{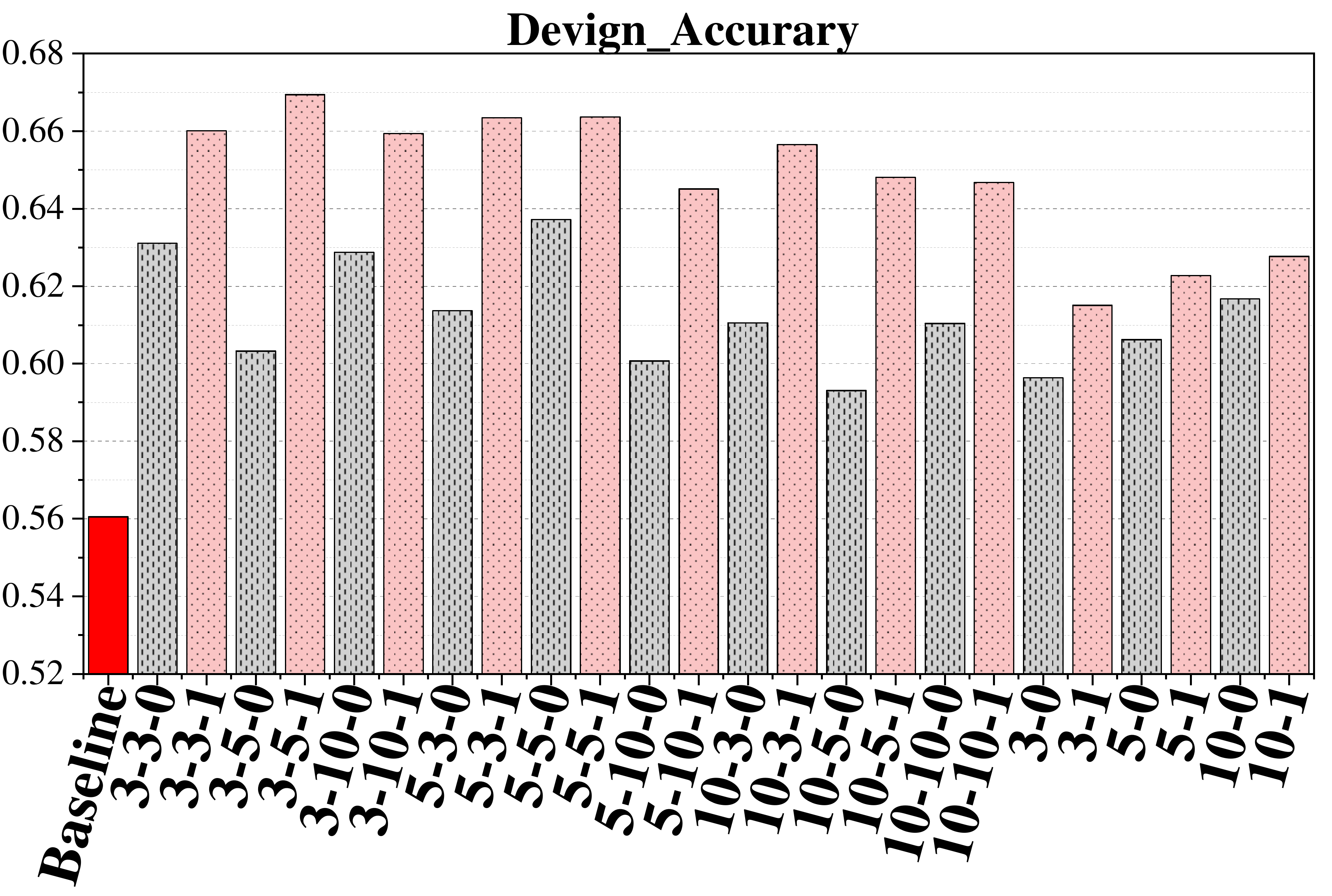}
\end{minipage}
}
\subfigure{
\begin{minipage}[t]{0.23\textwidth}
\centering
\includegraphics[width=\textwidth]{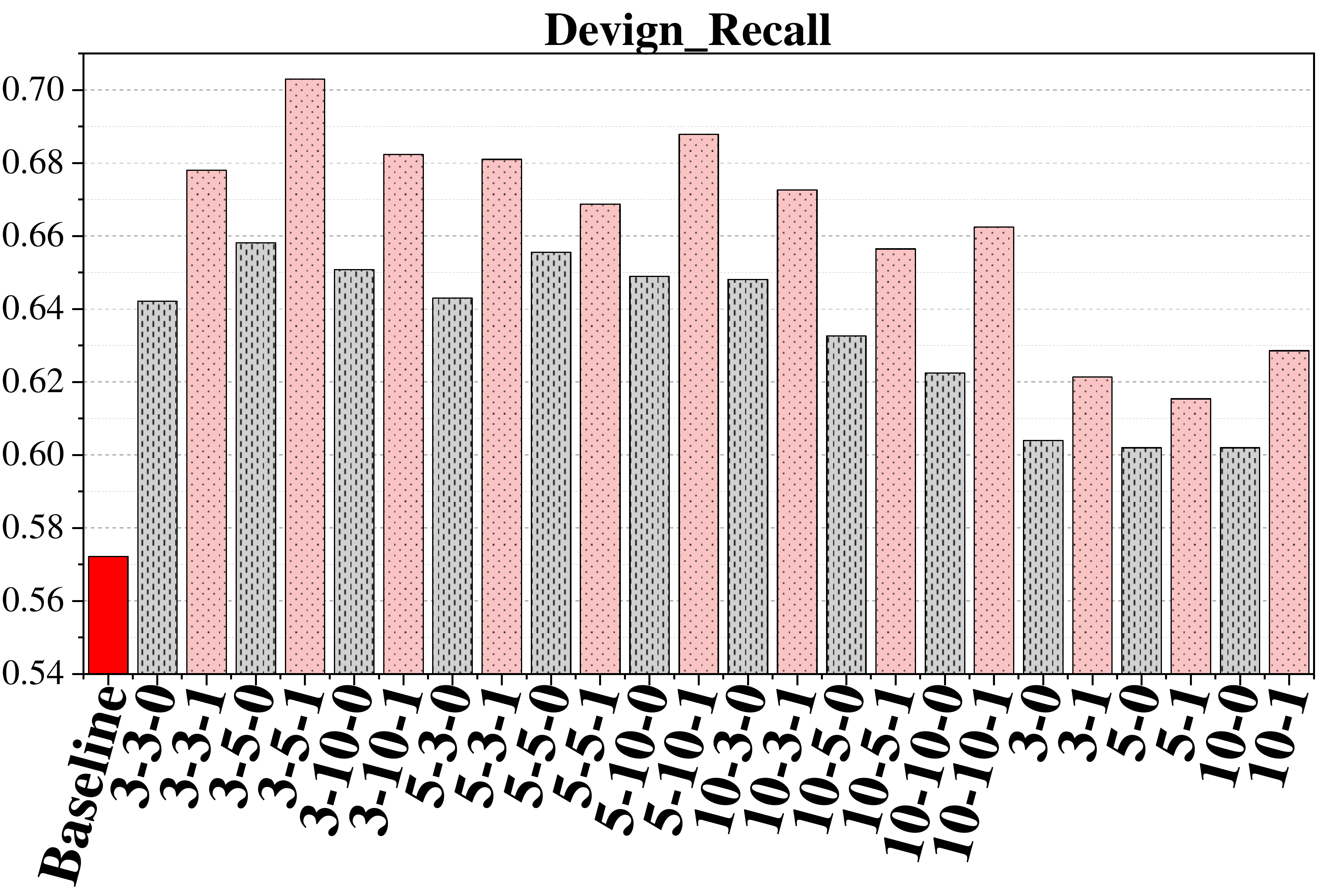}
\end{minipage}
}
\subfigure{
\begin{minipage}[t]{0.23\textwidth}
\centering
\includegraphics[width=\textwidth]{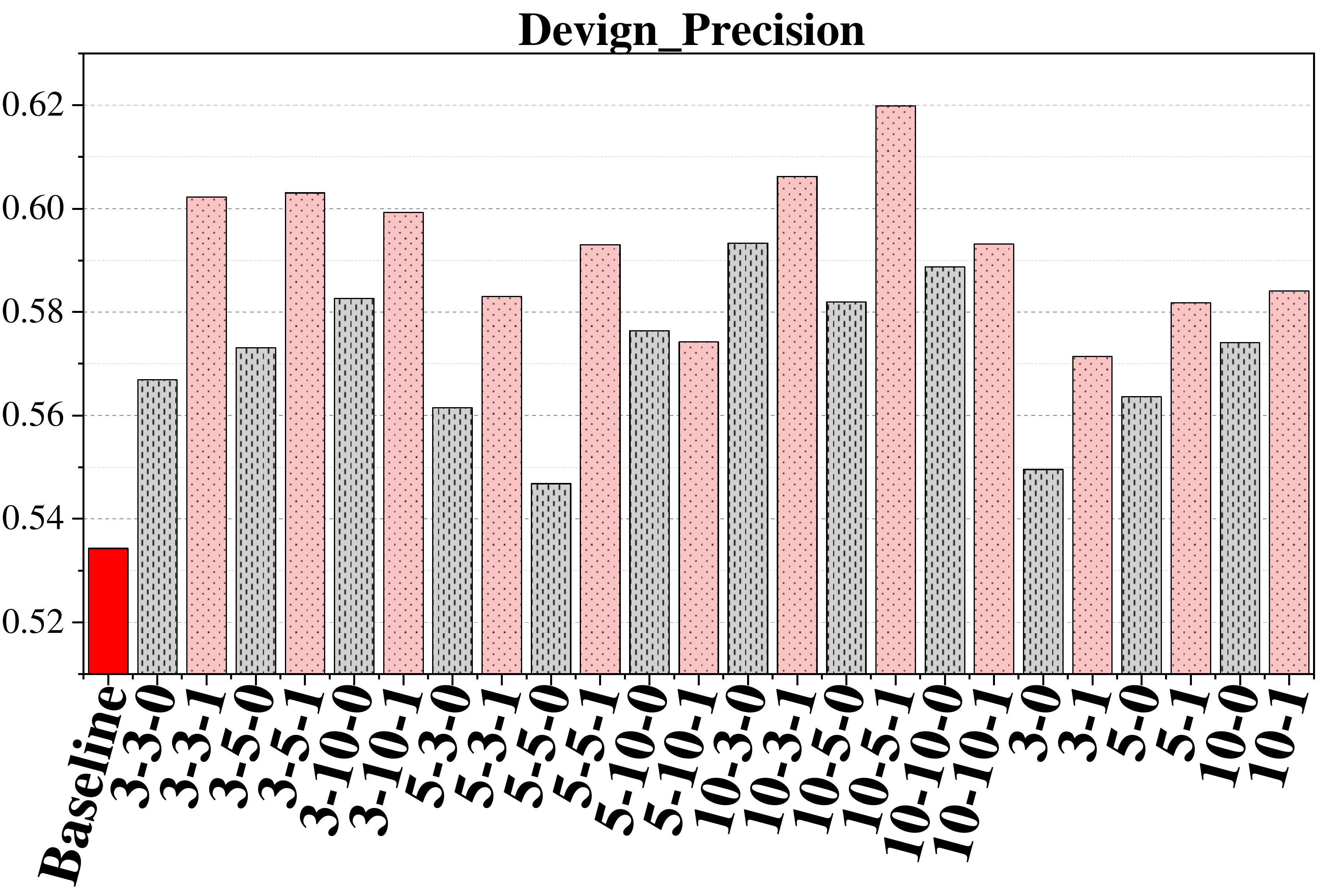}
\end{minipage}
}
\subfigure{
\begin{minipage}[t]{0.23\textwidth}
\centering
\includegraphics[width=\textwidth]{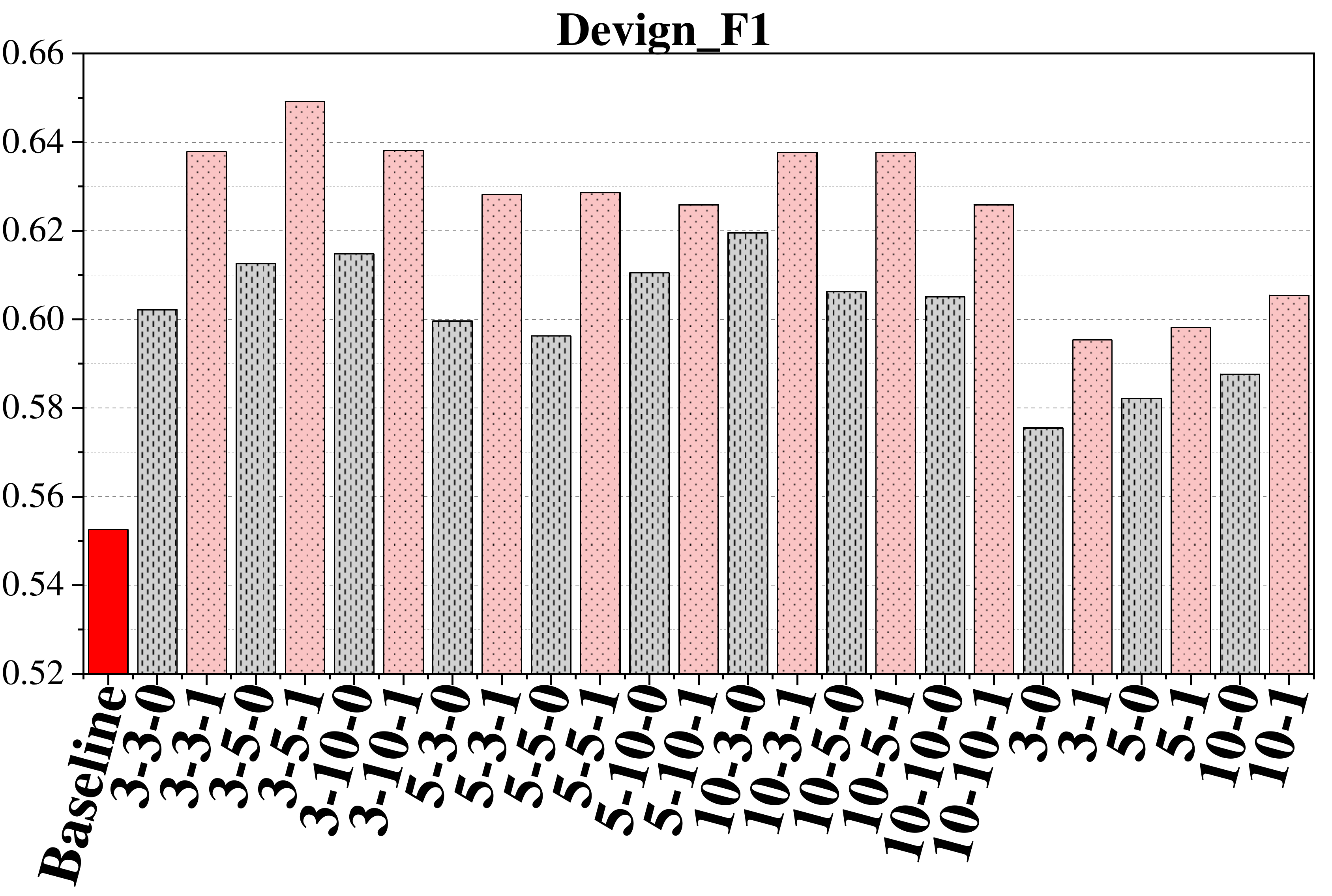}
\end{minipage}
}
\caption{Experimental results of \emph{Humer} with and without \emph{Data Augmenter}. 
The gray color shows the experimental result of \emph{Humer} without \emph{Data Augmenter}, and the pink color represents the experimental result of \emph{Humer} with \emph{Data Augmenter}.}
\label{fig:DA-results}
\end{figure*}

Obviously, for our tested DL-based vulnerability detection models (\ie \emph{TokenCNN}, \emph{VulDeePecker}, \emph{StatementGRU}, \emph{ASTGRU}, and \emph{Devign}), the detection ability can be improved after using \emph{Humer} to train them.
For example, the F1 is only 45.8\% when \emph{TokenCNN} is trained in a completely random order, but it can be increased to 57.3\% after applying \emph{Humer} to train the model.
Meanwhile, we also observe that when using \emph{Humer} to train different models, the improved effectiveness is different.
For instance, \emph{VulDeePecker}'s F1 can increase up to 10.9\% (62.8\%-51.9\%=10.9\%), while \emph{Devign}'s F1 can only increase up to 9.6\% (64.9\%-55.3\%=9.6\%).
It happens because different models may have different understandings of the same sample. 
This is similar to we humans, that is, different people may have different comprehension abilities when learning the same knowledge.

Table \ref{tab:CL-improvement} describes the improvement of \emph{TokenCNN}, \emph{VulDeePecker}, \emph{StatementGRU}, \emph{ASTGRU}, and \emph{Devign} after using \emph{Humer} to train them.
These results are obtained by computing the difference between \emph{Baseline} and \emph{Humer\_max} in Figure \ref{fig:CL-results}.
Through the table, we see that no matter which model, no matter which metrics, there are different degrees of improvement.
More specifically, using \emph{Humer} to train these five models can make them detect 12.6\% more vulnerabilities on average.
Meanwhile, the accuracy can also increase by an average of 9.95\% after using \emph{Humer}.
In other words, the use of \emph{Humer} can indeed enhance the detection ability of DL-based vulnerability detectors.

\subsection{For Data Augmenter}

In \emph{Humer}, we leverage \emph{Data Augmenter} to construct semantic-equivalent variants of those mispredicted samples which are used to fine-tune the model to strengthen the learning of ununderstood knowledge.
Therefore, in this part, we conduct comparative experiments to evaluate \emph{Humer} with and without \emph{Data Augmenter}.
Specifically, we construct an ablation experiment to achieve the goal, that is, only \emph{Data Augmenter} is deleted.
As shown in Figure \ref{fig:DA-results}, the red color represents the baseline result, the gray color represents the experimental result of \emph{Humer} without \emph{Data Augmenter}, and the pink color represents the experimental result of \emph{Humer} with \emph{Data Augmenter}.
To present the results more clearly, we use 0 and 1 to distinguish whether \emph{Data Augmenter} is removed in Figure \ref{fig:DA-results}.
For example, 3-5-0 means that the training set is first divided into three subsets to calculate the model-based difficulty. 
After sorting, the training set will be divided into five buckets, and then train a model by using \emph{Humer} without \emph{Data Augmenter}.
3-1 means that the training set is first sorted by their code complexity, and then is divided into three buckets to train a model by using \emph{Humer} with \emph{Data Augmenter}.

Through the results in Figure \ref{fig:DA-results}, we observe that whether \emph{Humer} has \emph{Data Augmenter} or not, after using \emph{Humer} for training, the detection effectiveness of DL-based vulnerability detection models can be improved.
But the overall effectiveness (\ie F1) will increase more when \emph{Humer} adopts \emph{Data Augmenter} to fine-tune the model.
In addition, we also find that there is a situation where after using \emph{Data Augmenter}, the recall of the model may be higher but the precision is slightly lower, or the precision becomes higher but the recall becomes a little lower.
For example, suppose that we first split the training set into five subsets to train five submodels for difficulty calculation, and then divide them into ten buckets for curriculum learning after sorting.
At this point, if we use \emph{Humer} without \emph{Data Augmenter} (\ie 5-10-0) to train \emph{Devign}, the recall and precision are 64.9\% and 57.6\%, respectively.
After adding \emph{Data Augmenter} into \emph{Humer} (\ie 5-10-1), the recall can increase to 68.8\% but the precision drops from 57.6\% to 57.4\%.

This phenomenon may correspond to the partial section phenomenon in human learning, that is, someone can do well in some subjects while poor in other ones.
Suppose there is a student Tom who is more interested in mathematics than literature, so he is willing to spend more time learning mathematics.
However, a person's energy is limited. 
Since what he has been studying is mathematics knowledge, he may have forgotten part of literary knowledge while gaining more mathematics knowledge.
Correspondingly, if there are many vulnerability samples in mispredicted data, most of the knowledge learned is the knowledge related to the vulnerability, which may cause the knowledge of some normal samples to be forgotten.
In the end, more vulnerabilities will be detected but more normal samples will be marked as vulnerabilities at the same time, that is, recall becomes higher but precision becomes lower.
To verify our guess, we compute the ratio of vulnerability samples to normal samples in \emph{Devign}'s mispredicted data (\ie 5-10-0 and 5-10-1).
After the analysis, we find that more than 90\% of the data are vulnerabilities, that is, the ratio of vulnerability samples to normal samples is more than 9:1.
But in fact, the overall detection ability (\ie F1) of these DL-based vulnerability detection models can be improved.
In other words, models can learn more knowledge after using \emph{Data Augmenter} than without using \emph{Data Augmenter}.

\subsection{Summary}

Through our experimental results, we obtain several findings:
\begin{enumerate}
\item For \emph{Humer}, model-based \emph{Difficulty Calculator} is better than code-based \emph{Difficulty Calculator}.
\item The training set is divided into different numbers of subsets (\ie $M$) and buckets (\ie $N$) can affect the detection effectiveness of \emph{Humer}.
\item After training with \emph{Humer}, the detection capabilities of DL-based vulnerability detection models can be improved.
\item For different DL-based vulnerability detection models, the enhancement effectiveness of using \emph{Humer} is different.
\item \emph{Humer} without \emph{Data Augmenter} can also enhance the detection ability of DL-based vulnerability detection models.
\item \emph{Humer} with \emph{Data Augmenter} is better than \emph{Humer} without \emph{Data Augmenter}.
\end{enumerate}

%% file: outline/casestudy.tex
\section{Case Study}

To further verify the practicability of \emph{Humer}, we conduct a case study to check whether the use of \emph{Humer} can help DL-based vulnerability detectors to discover new vulnerabilities from real-world open source software.
We select seven widely used open source products as our test objects:
\emph{Git} \cite{git}, \emph{Httpd} \cite{httpd}, \emph{Libav} \cite{libav}, \emph{OpenSSL} \cite{openssl}, \emph{Seamonkey} \cite{seamonkey}, \emph{Thunderbird} \cite{thunderbird}, and \emph{Xen} \cite{xen}.
The selected versions of these products include both several old versions and the latest version.
In this way, we can report whether vulnerabilities in old versions have been ``silently'' patched in the latest version or not.
As shown in Table \ref{tab:casestudy}, the total number of lines of code we analyzed exceeds 58 million lines. 

\begin{table}[htbp]
  \centering
  \small
  \caption{Summary of our tested open-source software}
    \begin{tabular}{|c|ccc|}
    \hline
    Software & \#Files & \#Functions & \#Lines of Code \\
    \hline
    Git & 1,427 & 24,384 & 844,049 \\
    Httpd & 1,338 & 16,398 & 765,375 \\
    Libav & 4,983  & 37,682  & 2,087,350  \\
    OpenSSL & 3,195 & 24,512 & 120,658  \\
    Seamonkey & 42,332 & 568,513 & 21,031,517 \\
    Thunderbird & 51,139 & 665,579 & 25,544,334 \\
    Xen & 14,829	& 207,447 &	8,279,792 \\
    \hline
    Total & 119,243 & 1,544,515 & 58,673,075 \\
    \hline
    \end{tabular}%
  \label{tab:casestudy}%
\end{table}%

\begin{table*}[htbp]
  \centering
  \footnotesize
  \caption{Vulnerabilities detected from Httpd product}
    \begin{tabular}{|c|c|c|c|c|}
    \hline
    Target product & CVE ID & Vulnerability release date & Vulnerable file in the target product & Patched or not in the latest version \\
    \hline
    \multirow{5}[0]{*}{Httpd-2.0.35} & CVE-2007-5000 & 12/13/2007 & modules/mappers/mod\_imap.c & deleted in Httpd-2.4.51 \\
          & CVE-2011-4317 & 11/29/2011 & modules/mappers/mod\_rewrite.c & patched in Httpd-2.4.51  \\
          & CVE-2013-5704 & 4/15/2014 & modules/loggers/mod\_log\_config.c & patched in Httpd-2.4.51  \\
          & CVE-2016-0718 & 5/26/2016 & srclib/apr-util/xml/expat/lib/xmltok\_impl.c       & deleted in Httpd-2.4.51 \\
          & CVE-2016-0718 & 5/26/2016 & srclib/apr-util/xml/expat/lib/xmlparse.c            & deleted in Httpd-2.4.51 \\
          \hline
    \multirow{7}[0]{*}{Httpd-2.2.14} & CVE-2009-3555 & 11/9/2009 & modules/ssl/ssl\_engine\_io.c & patched in Httpd-2.4.51  \\
          & CVE-2009-3555 & 11/9/2009 & modules/ssl/ssl\_engine\_init.c & patched in Httpd-2.4.51  \\
          & CVE-2011-4317 & 11/29/2011 & modules/mappers/mod\_rewrite.c & patched in Httpd-2.4.51  \\
          & CVE-2013-5704 & 4/15/2014 & modules/loggers/mod\_log\_config.c & patched in Httpd-2.4.51  \\
          & CVE-2014-0231 & 07/20/2014 & modules/generators/mod\_cgid.c & patched in Httpd-2.4.51  \\
          & CVE-2016-0718 & 5/26/2016 & srclib/apr-util/xml/expat/lib/xmltok\_impl.c       & deleted in Httpd-2.4.51 \\
          & CVE-2016-0718 & 5/26/2016 & srclib/apr-util/xml/expat/lib/xmlparse.c             & deleted in Httpd-2.4.51 \\
          \hline
    \multirow{2}[0]{*}{Httpd-2.4.51} & CVE-2014-0117 & 7/20/2014 & modules/proxy/proxy\_util.c & Not pacthed \\
          & CVE-2017-7668 & 6/19/2017 & server/util.c          & Not pacthed \\
          \hline
    \end{tabular}%

  \label{tab:httpd}%
\end{table*}%

To commence our case study, we first use data in Table \ref{tab:dataset} (\ie 12,460 vulnerabilities and 14,858 normal functions) to train five models (\ie \emph{TokenCNN} \cite{russell2018automated}, \emph{VulDeePecker} \cite{2018VulDeePecker}, \emph{StatementGRU} \cite{lin2019statementgru}, \emph{ASTGRU} \cite{feng2020astgru}, and \emph{Devign} \cite{zhou2019devign}) with \emph{Humer}.
After training, we then leverage these five enhanced models to scan vulnerabilities from products in Table \ref{tab:casestudy}.
Through the results in Table \ref{tab:DC-results}, we see that the detection performance of \emph{TokenCNN}, \emph{VulDeePecker}, \emph{StatementGRU}, \emph{ASTGRU}, and \emph{Devign} is not ideal when detecting real-world vunlerabilities.
Therefore, to make our scanning results more accurate, we only consider a function as a vulnerability and report it as a warning when all five models label it as vulnerable.
After scanning, we obtain a total of 343 warnings.
To check if they are real vulnerabilities or false positives, we conduct manual analysis to compare them with some vulnerabilities\footnote{We have collected some vulnerabilities with CVE-ID from NVD.} in NVD one by one. 
If it is found that the two belong to the same pattern, it is judged as a real vulnerability.
After analyzing all warnings, we find that 281 of them correspond to patterns of vulnerabilities in NVD, and the other 62 warnings are false positives.

Due to the limited page, we only present the scanning results from \emph{Httpd} product in this section, the other detected vulnerabilities can be seen on our website\footnote{https://github.com/Humer-DL/Humer.}.
Table \ref{tab:httpd} shows the details including the corresponding CVE ID in NVD, the release data in NVD, the vulnerable file in the target product, and whether patched or not in the latest version of the target product.
From the results in Table \ref{tab:httpd}, we find that some vulnerabilities can exist in multiple versions, and some vulnerabilities may be patched or deleted in the latest version after being discovered.
After statistical analysis, we observe that among 281 detected vulnerabilities, 98 of them have been patched by vendors in the latest version of corresponding products, 24 of them have been deleted, and the other 159 vulnerabilities still exist in the products.
We have reported these vulnerabilities to their vendors and hope that they can release a patched version as soon as possible.

\begin{figure}[htbp]
\centerline{\includegraphics[width=0.43\textwidth]{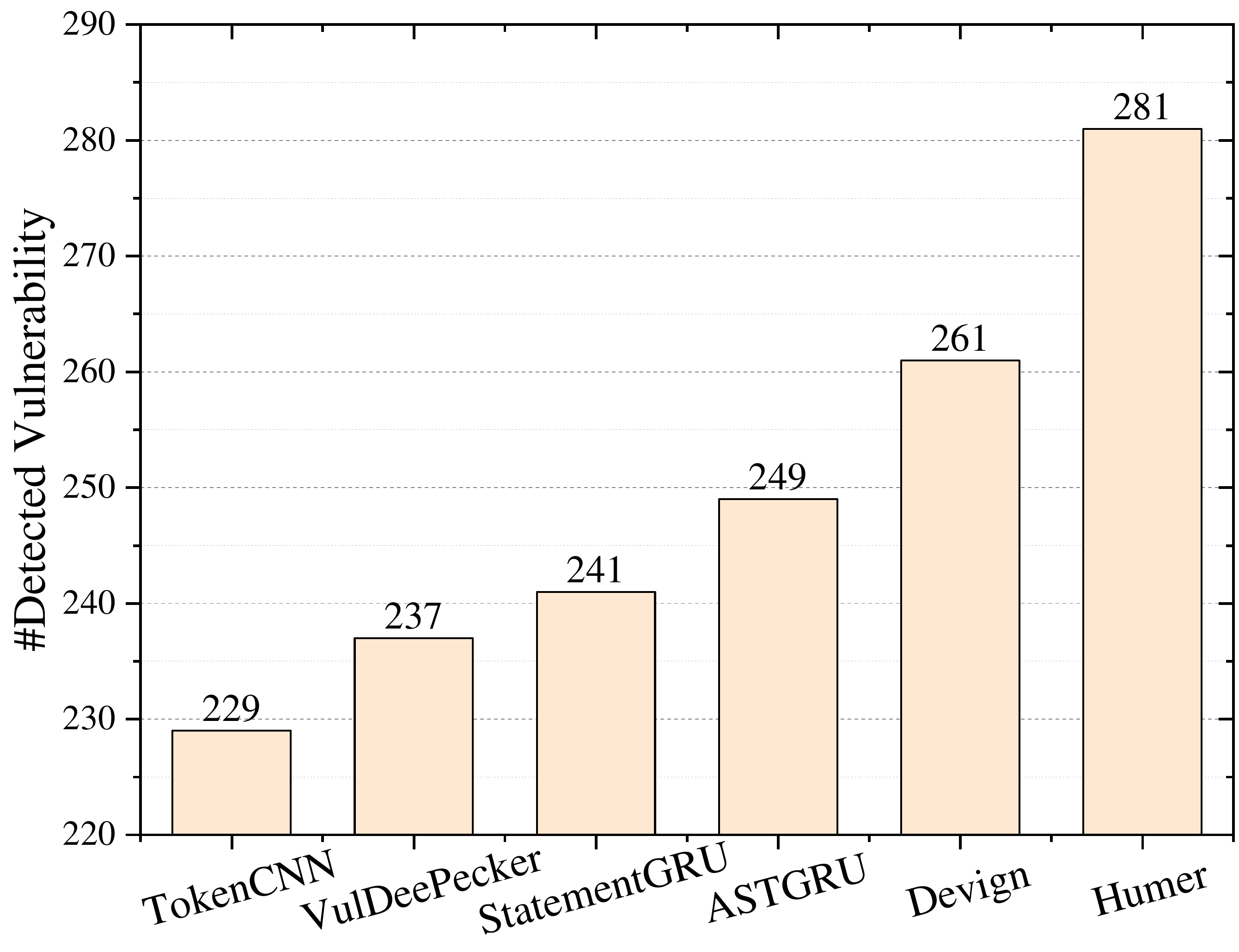}}
\caption{Detected results of 281 vulnerabilities scanned by DL-based vulnerability detectors (\ie \emph{TokenCNN}, \emph{VulDeePecker}, \emph{StatementGRU}, \emph{ASTGRU}, and \emph{Devign}) without \emph{Humer}}
\label{fig:casestudy}
\end{figure}

After collecting our detected vulnerability samples (\ie 281 vulnerabilities), we also conduct another experiment to check how many vulnerabilities cannot be detected by the baseline models, that is, vulnerabilities that can only be detected after using \emph{Humer} for training.
Figure \ref{fig:casestudy} presents the detected results of  \emph{TokenCNN}, \emph{VulDeePecker}, \emph{StatementGRU}, \emph{ASTGRU}, and \emph{Devign} without \emph{Humer}.
It shows that the use of \emph{Humer} can make DL-based vulnerability detection models detect more vulnerabilities.
In fact, after using \emph{Humer} to train \emph{TokenCNN}, \emph{VulDeePecker}, \emph{StatementGRU}, \emph{ASTGRU}, and \emph{Devign}, they can detect 52 (\ie 281-229=52), 44 (\ie 281-237=44),  40 (\ie 281-241=40), 32 (\ie 281-249=32), and 20 (\ie 281-261=20) more vulnerabilities, respectively.
In one word, using \emph{Humer} can enable them to detect 37 (\ie (52+44+40+32+20)/5=37.6) more vulnerabilities on average.
Such results demonstrate the effectiveness of \emph{Humer} on enhancing DL-based vulnerability detection models.

%% file: outline/discussion.tex
\section{Discussions}

As aforementioned, we select three different $M$ to split the training set into different subsets for model-based difficulty calculation and three different $N$ to divide the sorted training set into different buckets for curriculum learning.
In fact, although the number of $M$ and $N$ are both three, we conduct a total of 24 experiments for each DL-based vulnerability detection model (Figure \ref{fig:DA-results}).
In other words, the total number of experiments performed to evaluate \emph{Humer} is 120 (\ie 24$*$5=120) since we select five DL-based vulnerability detectors (\ie \emph{TokenCNN} \cite{russell2018automated}, \emph{VulDeePecker} \cite{2018VulDeePecker}, \emph{StatementGRU} \cite{lin2019statementgru}, \emph{ASTGRU} \cite{feng2020astgru}, and \emph{Devign} \cite{zhou2019devign}).
In our future work, we plan to buy more GPUs or spend more time to test more $M$ and $N$ to find more suitable results.

Because most of existing DL-based vulnerability detection methods are not open source, we only select five state-of-the-art models (\ie \emph{TokenCNN}, \emph{VulDeePecker}, \emph{StatementGRU}, \emph{ASTGRU}, and \emph{Devign}) to commence the evaluations of \emph{Humer}.
In fact, the code representations of these five methods are different, resulting in almost different neural networks used in the model.
In this way, we can achieve more comprehensive evaluations to examine the effectiveness of \emph{Humer} on enhancing DL-based vulnerability detection models.
In the future, we will choose more DL-based vulnerability detection approaches to evaluate \emph{Humer}.

From the results in Section IV.D, we see that in mispredicted data, if the ratio of vulnerability samples to normal samples is too different, the recall may increase but the precision may be slightly decreased.
But in fact, the F1 can be still increased, that is, the overall performance is still improved.
We plan to further analyze the semantic-equivalent variant samples to obtain the ratio of the vulnerability data to the normal data first.
If the amount of a certain type of data is small, we will input this type of data into \emph{Data Augmenter} and increase the number of code transformations and combinations to generate more of this type of data.
By this, we can balance the ratio of vulnerability samples to normal samples to mitigate the issue.

In \emph{Data Augmenter}, we design five code transformation rules to construct semantic-equivalent variant samples.
In fact, a program may satisfy several different rules.
In this paper, we propose two types of transformations to generate the variants, that is, a simple transformation and a hard transformation.
For simple transformation, we only apply one rule in one satisfied code fragment and other code remain unchanged.
For hard transformation, we apply all satisfied rules to generate a complex variant.
In our future work, we will implement more code transformation rules and design more different combinations of these rules to construct more semantic-equivalent variant samples of mispredicted data.

%% file: outline/relatedwork.tex
\section{Related Work}

\subsection{Vulnerability Detection}

Existing source code vulnerabilities detection methods can be classified into two main categories: the first is code-similarity-based approaches and the second is pattern-based techniques.

\subsubsection{Code-similarity-based Vulnerability Detection}

Code-similarity-based vulnerability detection methods apply different code similarity analysis algorithms to measure the similarity between known vulnerabilities and the program to be detected.
In fact, the similarity can be evaluated from different perspectives, such as string-based~\cite{jang2012redebug, kim2017vuddy}, tree-based~\cite{jiang2007deckard, pham2010detection}, token-based~\cite{ sajnani2016sourcerercc, kamiya2002ccfinder}, graph-based~\cite{li2012cbcd}, and their hybrid-based~\cite{li2016vulpecker}.
Because the similarity is measured by analyzing known vulnerabilities and the source code to be detected, these code-similarity-based vulnerability detection tools are used to find cloned vulnerabilities.

\subsubsection{Pattern-based Vulnerability Detection}

To discover more vulnerabilities, researchers propose to extract the patterns of vulnerabilities and detect new vulnerabilities by matching vulnerability patterns.
According to the degree of automation, these studies can be divided into three types.
The first type \cite{checkmarx, rats, flawfinder} is to define vulnerability patterns by human experts.
However, because experts cannot generate all patterns of different vulnerabilities, their detection effectiveness is not ideal.
The second type \cite{shankar2001detecting, backes2009automatic, yamaguchi2015automatic, shar2014web} is to extract certain features (\eg API symbols \cite{yamaguchi2012generalized}) first and then leverage some traditional machine learning algorithms (\eg support vector machine) to detect vulnerabilities.
The third type \cite{2018VulDeePecker, zou2019muvuldeepecker, zhou2019devign, lin2017poster, duan2019vulsniper, li2021sysevr, russell2018automated, wang2020funded, cheng2021deepwukong} is to use deep learning (DL) to detect vulnerabilities.

For example, \emph{VulDeePecker} \cite{2018VulDeePecker} first extracts the program slices of a program and then makes use of a \emph{bidirectional long short-term memory} (BLSTM) to train a detector.
\emph{muVulDeePecker} \cite{zou2019muvuldeepecker} improves the implementation of \emph{VulDeePecker}, it introduces the concept of code attention and uses it to help \emph{VulDeePecker} complete multiclass vulnerability detection. 
Feng \emph{et al.} design a tree-based vulnerability detector.
They first apply static analysis to extract the abstract syntactic tree of programs and then apply preorder traversal search algorithm to convert the trees into sequences.
Finally, these sequences will be used to train a \emph{bidirectional gated recurrent unit} (BGRU) model to detect vulnerabilities.
\emph{DeepWukong} \cite{cheng2021deepwukong} distills the program semantics into a program dependency graph and splits it into several subgraphs according to the program points of interest.
Then these subgraphs are fed into a \emph{graph neural network} (GNN) to train a vulnerability detector.

\subsection{Curriculum Learning}

Inspired by children's language learning, Elman \cite{elman1993learningCL2} studies the influence of learning programs on synthetic grammar tasks.
Through his experimental results, he finds that a recurrent neural network can learn a grammar when training data is presented from simple to complex order, but fails to do so when the order is random. 
In 2009, Bengio \emph{et al.} \cite{bengio2009curriculum} officially name this learning method as \emph{curriculum learning} (CL) and verify that CL can not only improve the effectiveness but also speed the convergence of the model.
Since then, CL has been widely used in various fields \cite{florensa2017reverseCL001, el2020studentCL002, guo2018curriculumnetCL003, tay2019simpleCL004, zhou2020uncertaintyCL005}.

For example, 
El-Bouri \emph{et al.} \cite{el2020studentCL002} combine CL with reinforcement learning to predict where in a hospital emergency patients will be admitted after being triaged. 
Guo \emph{et al.} \cite{guo2018curriculumnetCL003} use CL to help train deep neural networks on large-scale weakly-supervised web images. 
Tay \emph{et al.} \cite{tay2019simpleCL004} introduce CL to tackle the problem of reading comprehension over long narratives. 
Zhou \emph{et al.} \cite{zhou2020uncertaintyCL005} use uncertainty-aware CL in neural machine translation to improve the translation quality and convergence speed. 
Zhao \emph{et al.} \cite{zhao2021automaticCL006} propose a CL-based Deep QNetwork and introduce a teacher policy model to replace the traditional random sampling method, which significantly improves the effectiveness and stability of the dialogue tasks. 
Dai \emph{et al.} \cite{dai2021previewCL007} make use of a curriculum structure and a schema structure to improve the effectiveness of dialog state tracking.
More related work can be found in a recent survey \cite{wang2021survey}.

%% file: outline/conclusion.tex
\section{Conclusion}

In this paper, we propose to train DL-based vulnerability detection models in a human-learning manner.
Specifically, we implement a model-agnostic framework namely \emph{Humer}, and select five state-of-the-art vulnerability detection methods (\ie \emph{TokenCNN} \cite{russell2018automated}, \emph{VulDeePecker} \cite{2018VulDeePecker}, \emph{StatementGRU} \cite{lin2019statementgru}, \emph{ASTGRU} \cite{feng2020astgru}, and \emph{Devign} \cite{zhou2019devign}) to evaluate \emph{Humer} on a real-world vulnerability dataset.
The experimental results indicate that using \emph{Humer} can make them detect 12.6\% more vulnerabilities on average.
We also conduct a case study to demonstrate the practicability of \emph{Humer} and finally discover 281 vulnerabilities that are not reported in NVD.